%% file: experiment.tex
\documentclass{svmult}

\usepackage{makeidx}         % allows index generation
\usepackage{graphicx}        % standard LaTeX graphics tool
\usepackage{natbib}
\usepackage{amsmath,amssymb}
\usepackage{nomencl}
\let\abbrev\nomenclature

\usepackage{url}

\usepackage{multicol}        % used for the two-column index
\usepackage[bottom]{footmisc}% places footnotes at page bottom
\makeindex             % used for the subject index
\def\beq{\begin{equation}}
\def\eeq{\end{equation}}
\def\beqn{\begin{eqnarray}}
\def\twiddle{\lower.9ex\rlap{$\kern-.1em\scriptstyle\sim$}}
\def\bigtwiddle{\lower1.ex\rlap{$\sim$}}
\def\gtwid{\mathrel{\raise.3ex\hbox{$>$\kern-1.05em\lower1ex\hbox{
$\sim$}}}}
\def\ltwid{\mathrel{\raise.3ex\hbox{$<$\kern-1.05em\lower1ex\hbox{
$\sim$}}}}

\newcommand{\be}{B_\oplus}
\newcommand{\re}{R_\oplus}
\def\MET{\mbox{${\hbox{$E$\kern-0.6em\lower-.1ex\hbox{/}}}_T$}}

\begin{document}

\title*{Axion Searches in the Past, at Present, and in the Near Future}
\label{chap:experiments}

\author{R\'{e}my Battesti\inst{1}\and Berta Beltr\'{a}n\inst{2}\and Hooman
  Davoudiasl\inst{3}\and Markus Kuster\inst{4,5,*}\and Pierre
  Pugnat\inst{6}\and Raoul Rabadan\inst{7}\and Andreas Ringwald\inst{8}\and
  Neil Spooner\inst{9}\and Konstantin Zioutas\inst{6,10}}

\authorrunning{R. Battesti et al.}
\titlerunning{Experimental Axion Searches}

\institute{ 
  Laboratoire National des Champs Magn\'etiques Puls\'es,
  CNRS/INSA/UPS, UMR 5147, 31400 Toulouse, France 
  \and 
  Department of Physics, Queen's University, Kingston, Ontario K7L 3N6,
  Canada 
  \and 
  Department of Physics, University of Wisconsin, Madison, WI 53706, USA
  \and 
  Technische Universit\"at Darmstadt, IKP, Schlossgartenstrasse~9, D-64289
  Darmstadt, Germany 
  \and Max-Planck-Institut f\"ur extraterrestrische Physik,
  Giessenbachstrasse, D-85748 Garching, Germany
  \and 
  European Organization for Nuclear Research (CERN), CH-1211 Gen\`eve 23,
  Switzerland
  \and 
  Institute for Advanced Study, Einstein Drive,
  Princeton, NJ, 08540, USA
  \and 
  Deutsches Elektronen-Synchrotron DESY, Notkestr. 85, D-22607 Hamburg,
  Germany
  \and 
  University of Sheffield, Sheffield S3 7RH, UK 
  \and 
  University of Patras, Rio, 26500 Patras, Greece\\ 
%  \texttt{battesti@lncmp.org},
%   \texttt{berta.beltran@cern.ch},
%   \texttt{hooman@hep.wisc.edu}, 
  \texttt{$^*$markus.kuster@cern.ch}
%   \texttt{Pierre.Pugnat@cern.ch}, 
%   \texttt{rabadan@ias.edu},
%   \texttt{andreas.ringwald@desy.de}, 
%   \texttt{n.spooner@shef.ac.uk},
%   \texttt{konstantin.zioutas@cern.ch}
 }

\maketitle
\begin{abstract}
  Theoretical axion models state that axions are very weakly interacting
  particles. In order to experimentally detect them, the use of colorful
  and inspired techniques becomes mandatory. There is a wide variety of
  experimental approaches that were developed during the last 30 years,
  most of them make use of the Primakoff effect, by which axions convert
  into photons in the presence of an electromagnetic field. We review the
  experimental techniques used to search for axions and will give an
  outlook on experiments planned for the near future.
\end{abstract}

\section{Searches for Dark Matter Axions}
\begin{figure}[t]
  \centering
  \includegraphics[width=0.8\textwidth]{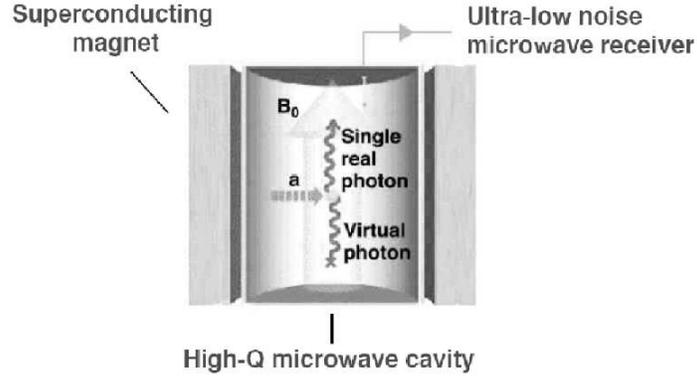}
  \caption{Schematic principle of the microwave cavity experiment to 
    look for cold dark matter axions. First an axion would be resonantly
    converted into a quasi-monochromatic photon in the magnetic-field
    permeated microwave cavity. Then an ultra-low-noise receiver is set to
    detect this photon as an excess on the cavity power output}
  \label{fig:ex-microcavity}
\end{figure}

As described in the review by P.~Sikivie on cosmological axions in Chap.
\ref{chap:axion-cosmology}, axions produced in the early universe with a
mass ranging from $\umu\text{eV}$ to $\text{meV}$ could account for the
cold dark matter component of the universe and therefore an experimental
evidence for axions in this mass range would be of great interest nowadays.
Microwave resonant cavity experiments are the most sensitive detectors for
cold-dark matter axions at present. The heart of such a device is a
metallic conductor which acts as a cavity for electromagnetic fields such
as photons. By tuning the size of the magnetic-field permeated cavity its
resonant frequency can be shifted such that it matches the frequency of the
axion field (which is related to the axion mass). In this case axions will
convert resonantly into quasi-monochromatic photons which could be detected
with sensitive microwave receivers as an excess on the cavity power output.
In Fig.  \ref{fig:ex-microcavity} the schematic working principle of this
kind of detectors is shown.

\begin{figure}[t]
  \centering
  \includegraphics[width=0.7\textwidth]{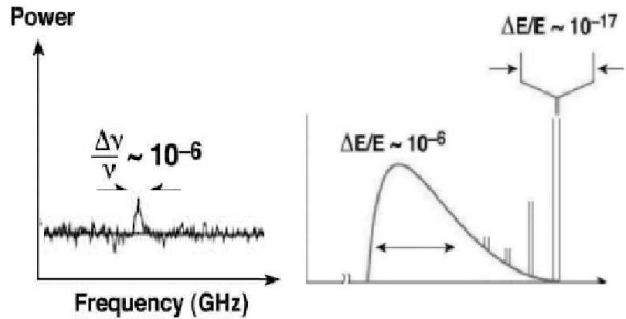}
  \caption{Left: Representation of the signal expected in a microwave cavity
    experiment, where the photon signal appears as an excess over the noise
    in the power frequency spectrum. The width of the signal would be
    related with the predicted axion halo velocities.  Right: Zoom showing
    the possible fine structure details in the expected signal}
  \label{fig:ex-2microcavity}
\end{figure}

The signal obtained would carry information not only about the axion mass,
but also about the axion distribution in the galactic halo, as its width
would represent the virial distribution of thermalized axions.
Furthermore, the signal may also posses a finer substructure due to axions
that have recently fallen into the galaxy and have not yet thermalized, as
shown in Fig. \ref{fig:ex-2microcavity}. Again the reader is refered to
Skivie's review for a more detailed description of the axion distribution
in the galactic halo.

Due to the very low mass range of the dark matter axion, the expected
signal is minuscule and therefore the sensitivity of the experiment
crucially depends on the quality of the microwave receivers (see Chap.
\ref{chap:microwave-cavity} by Carosi and van Bibber). In the 80s two
pioneering experiments were carried out at Rochester-Brookhaven-Fermilab
(RBF) \cite{ex-DePanfilis:1987dk} and the University of Florida (UF)
\cite{ex-Hagmann:1990tj} using small volume cavities ($\sim 1\,\text{l}$)
and HFET amplifiers, but their sensitivities
were lacking by two or three orders of magnitude the sensitivity necessary
to probe the theoretically motivated region in the axion parametric space.

An improvement in the sensitivity lead to the second generation experiments
such as ADMX \cite{ex-Asztalos:2003px} which is currently taking data with
a bigger resonant cavity as the precedent experiment, but still making use
of the HFET amplifiers. Their results are compatible with the absence of
any axion signal, providing the best upper limits for the axion to photon
coupling constant $g_{a\gamma\gamma}$ in the lowest mass range ($m_a
\approx 10^{-4}-10^{-6}\,\text{eV}$, see Fig. \ref{fig:ex-exclusion-plot}).
A new amplifier technology based on Superconducting Quantum Interference
Devices (SQUID), currently under development, can further improve the
sensitivity of ADMX, such that it is possible to probe the theoretically
favoured axion parameter range ( see Chap.  \ref{chap:microwave-cavity} and
Fig. \ref{fig:ex-exclusion-plot}).

Another second generation experiment is the CARRACK experiment in Kyoto
\cite{ex-Tada:1999tu}.  They also use a resonant cavity to produce the
axion-converted-photons, but the signal detection is based on a Rydberg
atom single-quantum detector, providing an extremely low-noise photon
counting capability. This experiment is still in the development phase.

\section{Solar Axions Searches}
\begin{figure}[t]
  \centering
  \begin{minipage}{0.49\textwidth}
    \includegraphics[width=0.98\textwidth]{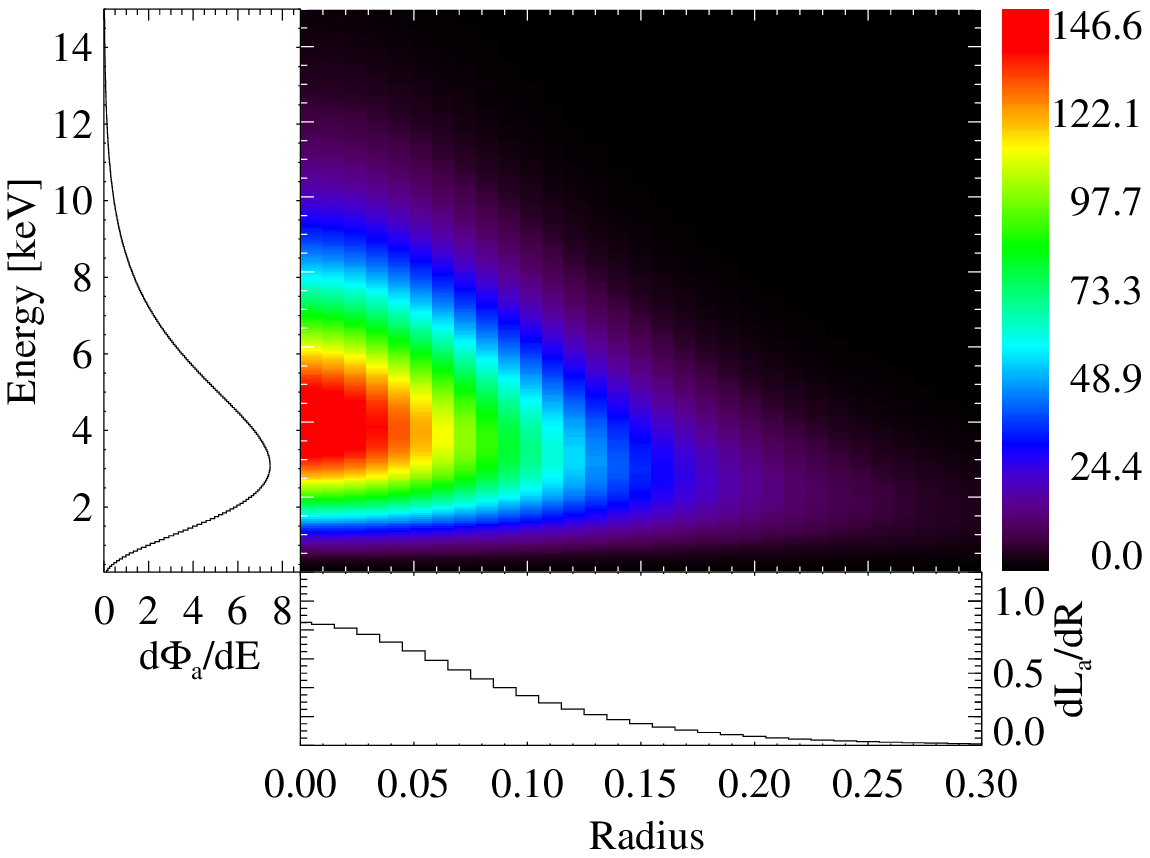}
  \end{minipage}
  \hfill
  \begin{minipage}{0.49\textwidth}
    \includegraphics[width=0.98\textwidth]{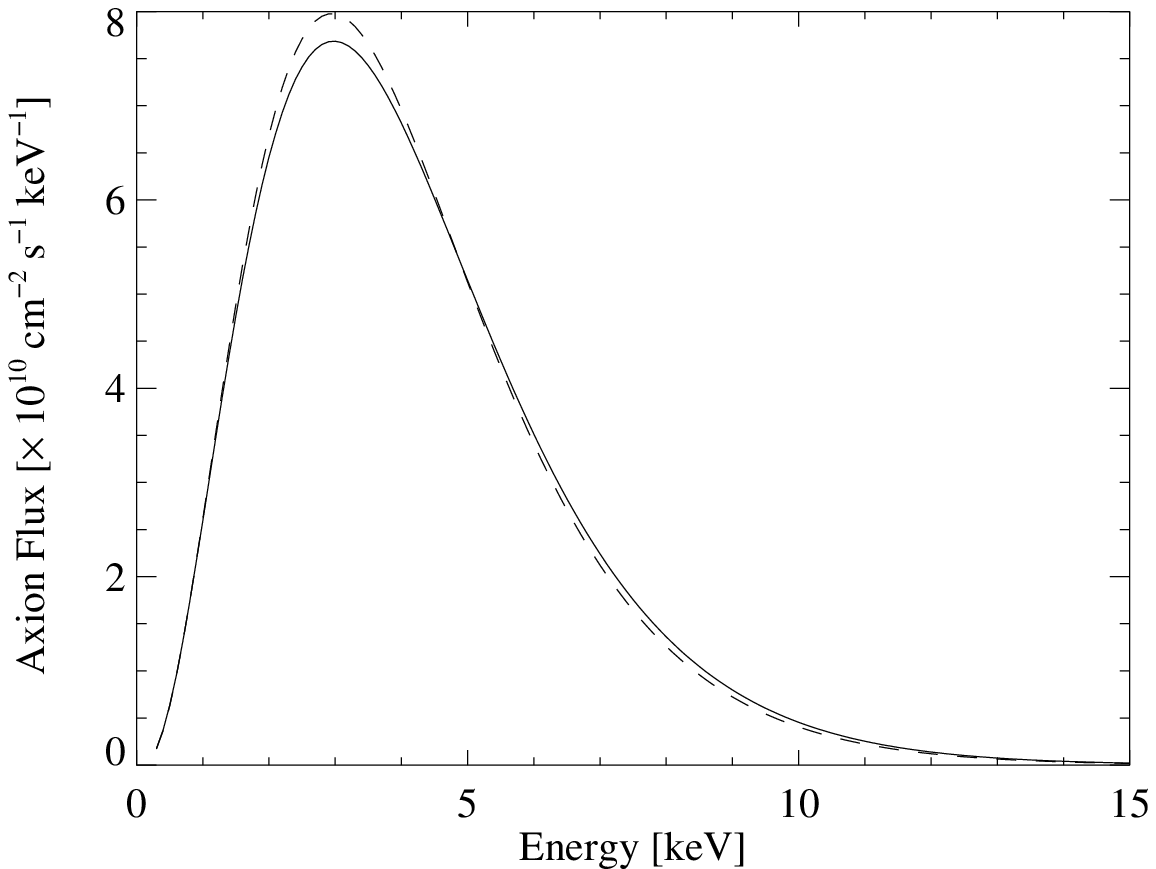}
  \end{minipage}
  \caption{Left: 2D representation of the axion surface luminosity
    seen from the Earth~\cite{ex-Andriamonje:2007ew} as a function of the
    solar radius and the axion energy.\index{Axion!Solar!Surface
      luminosity} Right: Comparison of the solar axion flux calculated
    using an older version of the standard solar model from 1982 (dashed
    line) and a more recent version from 2004 one (solid line)
    \cite{ex-bahcall:04a}.  Here an axion-photon coupling constant of $1
    \times 10^{-10}\, {\rm GeV}^{-1}$ is assumed}
  \label{fig:ex-axionflux}
\end{figure}

As seen from the Earth, the most important and strongest astrophysical
source for axions is the core of the Sun. There, pseudoscalar particles
like axions would be continuously produced in the fluctuating electric and
magnetic fields of the plasma via their coupling to two photons. After
production the axions would freely stream out of the Sun without any
further interaction (given that the axion mean free path in the Sun is
$\approx g_{10}^{-2}\,6\times10^{24}\,\text{cm}$ for $4\,\text{keV}$
axions~\footnote{Here we define
  $g_{10}=g_{a\gamma\gamma}/10^{-10}\,\text{GeV}^{-1}$.}, this is a valid
assumption). The resulting differential solar axion flux on Earth would be
\cite{ex-Andriamonje:2007ew}\index{Axion!Solar!Spectrum}
\begin{equation}
  \label{eq:ex-solar-axion-spectrum}
  \frac{\D\Phi_a}{\D E} = 6.02 \times
  10^{10}\,\text{cm}^{-2}\,\text{s}^{-1}\,\text{keV}^{-1}g_{10}^2\,
  E^{2.481}e^{-E/1.205}\;.
\end{equation}
The spectral energy distribution of the axions follows a thermal energy
distribution between $1$ and $10\,\text{keV}$, which peaks at $\approx
3\,\text{keV}$ as shown in the right part of Fig. \ref{fig:ex-axionflux}.
For comparison two different solar axion spectra are shown in Fig.
\ref{fig:ex-axionflux}, calculated from different versions of the standard
solar model (SSM)\abbrev{SSM}{Standard Solar Model}: the solid line
corresponds to a more recent implementation of the SSM from 2004, while the
spectrum plotted as a dashed line is based on an older version published in
1982 \cite{ex-Andriamonje:2007ew,ex-bahcall:04a}. It is evident that small
changes of the underlying SSM have only a small influence on the axion flux
and the corresponding shape of the axion energy spectrum.  In the following
we will summarize different techniques which aim to directly detect ``solar
axions''.

\subsection{Axion Helioscope}
\begin{figure}[t]
  \centering
  \includegraphics[width=0.8\textwidth]{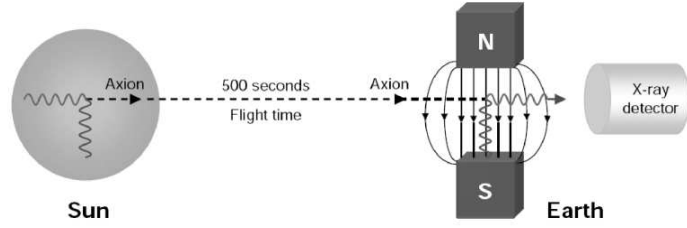}
  \caption{Working principle of an axion helioscope. Axions produced in the 
    Sun core by the Primakoff effect would be converted back into photons
    in a strong magnetic field. Eventually these photons, which have the
    same energy spectrum as the incoming axions, could be collected by a
    X-ray detector placed a the end of the magnet field area}
  \label{fig:ex-work}
\end{figure}

\index{Helioscope search|(} The most sensitive axion experiments at present
in the mass range of $10^{-5}\,\text{eV}\lesssim m_{a} \lesssim
1\,\text{eV}$ are so called ``axion helioscopes'', i.e., magnetic solar
telescopes. The underlying physical principle of such a telescope is based
on the idea proposed by P.~Sikivie in 1983 \cite{ex-Sikivie:1983ip} and is
illustrated in Fig.  \ref{fig:ex-work}. If we assume that axions are
produced in the Sun, they would reach the Earth after about $500\,\text{s}$
as an almost parallel axion beam (the apparent diameter of the axion
emission region on the solar disk is $\approx 0.1^\circ$).  The apparatus
for the detection on Earth uses the time reversed Primakoff effect, i.e.,
$a + \gamma_{\text{virtual}}\rightarrow\gamma$, where the axion interacts
with a virtual photon provided by a transversal magnetic field and
reconverts into a real photon. These ``reconverted'' photons have the same
energy and momentum as the axion, and therefore their energy distribution
is identical to the energy spectrum of the solar axions shown in Fig.
\ref{fig:ex-axionflux}.  An X-ray detector placed on the end side of the
magnetic field would collect the photons and search for an axion signature
above detector background. The expected number of reconverted photons
reaching the detector for the given differential solar axion spectrum is
\begin{equation}
  N=\int\frac{\D\Phi_a}{\D E}P_{a\to\gamma}\,S\,t\,\D E\;.
  \label{eq:ex-n-of-photons}
\end{equation}
Here $d\Phi_a/dE$ is the differential axion flux at the Earth, $S$ the
magnet bore area, $t$ the measurement time, and $P_{a\to\gamma}$ the
conversion probability of an axion into a photon. To maximize $t$ the
magnet is mounted in a similar way as conventional telescopes, such that it
can follow the track of the Sun on the sky.

The number of photons leaving the magnetic field towards the detector is
determined by the probability that an axion converts back to a
``observable'' photon inside the magnetic field, assuming that the
conversion volume (magnetic field area) is evacuated, the probability is
given by\index{Helioscope search!Conversion probability}
\begin{equation}
  \label{eq:ex-probability}
  P_{a\to\gamma}=\left(\frac{Bg_{a\gamma\gamma}}{2}\right)^2\,2L^2\,\frac{1-\cos(qL)}{(qL)^2}\;, 
\end{equation}
where $B$ and $L$ are the magnet field strength and length of the field,
and\index{Momentum transfer|\\see Helioscope }\index{Helioscope search!Coherence|(}
\begin{equation}
  \label{eq:ex-momentum-transfer}
  q=m^2_a/2E_\gamma\;,
\end{equation}
\index{Helioscope search!Momentum transfer}is the longitudinal momentum
difference (or momentum transfer) between the axion and the photon of the
energy $E_\gamma$. In order to achieve a maximum conversion probability
$P_{a\to\gamma}$ (constructive interference), the axion and photon fields
must remain in phase over the length of the magnetic field, this sets the
requirement that $qL<\pi$ \cite{ex-Lazarus:1992ry,ex-vanBibber:1988ge}.
Thus, the sensitivity of a helioscope is limited to a specific axion mass
range for a given $q$.  For increasing axion masses the conversion is
suppressed by the momentum mismatch between the axion and the photon,
namely due to the form factor $(1-\cos(qL))/(qL)^2$ in
\eqref{eq:ex-probability}.

Coherence can be restored for higher axion masses by filling the magnetic
conversion region with a buffer gas \cite{ex-vanBibber:1988ge}, such that
the photons inside the magnet pipe acquire an ``effective mass'' (which is
equivalent to a refractive index $n$). Consequently the wavelength of the
photon can match that of the axion for an appropriate gas pressure and
coherence will be preserved for a narrow axion mass window. In this case
the transition probability $P_{a\to\gamma}$ has to be recast into the more
general equation \cite{ex-vanBibber:1988ge}\index{Helioscope search!Buffer
  gas}\index{Helioscope search!Conversion probability}
\begin{equation}
  \label{eq:ex-prob-gas}
  P_{a\to\gamma}=\left(\frac{Bg_{a\gamma\gamma}}{2}\right)^2
    \frac{1}{q^2+\Gamma^2/4}[1+e^{-\Gamma 
    L}-2e^{-\Gamma L/2}\cos(qL)]\;.
\end{equation}
Here $\Gamma$ is the absorption coefficient for the X-rays in the medium
and the momentum transfer becomes
\begin{equation}
  \label{eq:ex-momentum2}
  q=\left| \frac{m_{\gamma}^2-m_a^2}{2E_a} \right|\;,
\end{equation}
with the effective photon mass $m_{\gamma}$\index{Effective photon mass}
\begin{equation}
  \label{eq:ex-buffer}
  m_\gamma\simeq \sqrt{\frac{4\pi \alpha  n_e}{m_e}}=28.9
  \sqrt{\frac{Z}{A}\varrho}\;, 
\end{equation}
where $n_e$ is the number density of electrons in the medium and $m_e$ the
electron mass. The second expression in \eqref{eq:ex-buffer} is a more
convenient representation for $m_{\gamma}$ giving the dependence of
$m_{\gamma}$ on the atomic number $Z$ of the buffer gas, its mass $A$, and
its density $\varrho$ in $\text{g}\,\text{cm}^{-3}$. Since Helium is the
type of gas commonly used for this purpose \eqref{eq:ex-buffer} can be
further simplified to
\begin{equation}
  \label{eq:ex-buffer2}
  m_\gamma (\text{eV})\simeq \sqrt{0.02\frac{P(\text{mbar})}{T(\text{K})}}\;,
\end{equation}
when we use the corresponding parameters $A$, $Z$, and $\varrho$ for Helium
and treat Helium as an ideal gas. This equation directly relates $m_\gamma$
to the measured experimental parameters, i.e., the pressure $P$ and
temperature $T$ of the gas inside the magnetic field region. For a specific
pressure $P$ and temperature $T$, the coherence is restored for a narrow
axion mass window, for which the effective mass of the photon matches that
of the axion such that
\begin{equation}
  \label{eq:ex-coherence2}
  qL < \pi\; \Longrightarrow \; \sqrt{m_\gamma^2-\frac{2\pi
      E_a}{L}}<m_a<\sqrt{m_\gamma^2+\frac{2\pi E_a}{L}}\;.
\end{equation}\index{Helioscope search!Coherence|)}The differential energy
spectrum of the photons from axion conversion follows from
\eqref{eq:ex-solar-axion-spectrum} and \eqref{eq:ex-probability}. Assuming
full coherence it can be approximated analytically by
\cite{ex-Andriamonje:2007ew}
\begin{eqnarray}
  \label{eq:ex-diff-axion-photon-spectrum}
  \frac{\D \Phi_\gamma}{\D E} &=& \frac{\D \Phi_a}{\D E} P_{a\to\gamma} \nonumber \\
  &=& 0.088\,\text{cm}^{-2}\,\text{day}^{-1}\,\text{keV}^{-1}
  g_{10}^4
  E^{2.481}e^{-E/1.205}\left(\frac{BL}{9.0\,\text{T}\,9.26\,\text{m}}\right)^2\,.
\end{eqnarray}
It is obvious that the resulting low photon count rates from
\eqref{eq:ex-diff-axion-photon-spectrum} require background optimized
detectors in order to achieve a reasonable detection sensitivity with an
axion helioscope.

\paragraph{Helioscope Implementations}
\index{Axion!Limits|(} The first axion helioscope was built at the
beginning of the 90s \cite{ex-Lazarus:1992ry}, using a $180\,\text{cm}$
long magnet powered with a $2.2\,\text{T}$ transverse magnetic field. As a
detector a proportional counter operating with a gas mixture of argon and
methane ($90\%$ to $10\%$) was used to collect the photons from axion to
photon conversion. The helioscope could be pointed to the Sun for a period
of $\approx 15\,\text{min}\,\text{day}^{-1}$. Two different regions of the
axion mass range (with and without a gas in the conversion volume) were
explored with this setup, setting the following limits on
$g_{a\gamma\gamma}$ and $m_a$ in the absence of a signal
\begin{equation}
  \begin{split}
    \label{eq:ex-lazarus}
    &g_{a\gamma\gamma} < 3.6 \times 10^{-9}\; {\rm GeV}^{-1}\;\;\;{\rm for\;\;}m_a < 0.03\;
    {\rm eV}\;,\\
    &g_{a\gamma\gamma} < 7.7 \times 10^{-9}\; {\rm GeV}^{-1}\;\;\;{\rm for\;\;}0.03\;
    {\rm eV} <m_a < 0.11 \;{\rm eV}\;.
  \end{split}
\end{equation}
\index{Helioscope search!Tokyo}\index{Tokyo helioscope} In the late 90s a
new helioscope with improved sensitivity ($2.3\,\text{m}$ long magnet,
$4\,\text{T}$) was built in Tokyo~\cite{ex-Moriyama:1998kd}.  For this
experiment sixteen PIN photodiodes were used as X-ray detectors and the
superconducting magnet was fixed on a polar mount, allowing the system to
follow the Sun in a range from $-28^{\circ}$ to $+28^{\circ}$ in
declination and $360^\circ$ in right ascension. Due to the better tracking
system, the Tokyo helioscope was able to follow the Sun for $24\,\text{h}$
a day and the observation time of the Sun could be significantly improved.
In total the Tokyo collaboration was able to use about 50\% of the duty
cycle of their instrument to track the Sun, while the rest of the time was
devoted to background measurements.  Similar to the first approach of
\cite{ex-Lazarus:1992ry}, the experiment was divided into two phases, the
first phase was finished in 1997 with an evacuated conversion region and
the second phase in 2000 when the magnet pipe was filled with helium gas in
order to extend the sensitive axion mass range.  The resulting upper limits
on $g_{a\gamma\gamma}$ from both data taking periods are
\cite{ex-Inoue:2000bj,ex-inoue:2002qy}
\begin{figure}[t]
  \centering
  \includegraphics[width=0.98\textwidth]{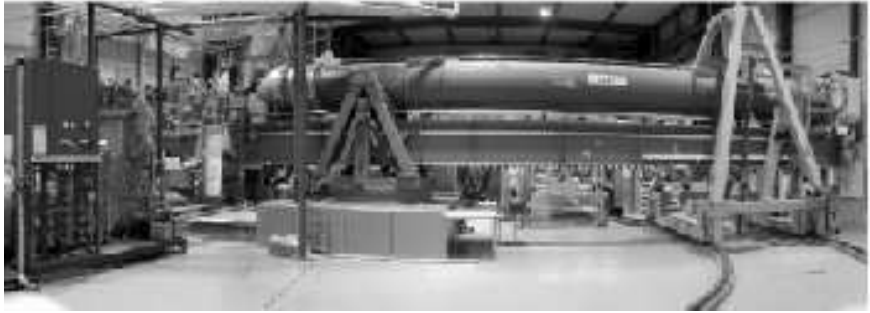}
  \caption{Experimental setup of the CAST experiment at CERN. From the left
    to the right the Helium cryogenic system, the CAST superconducting
    magnet, and the tracking system is shown. The tracking system allows to
    follow the Sun $1.5\,\text{h}$ per day during sunrise and sunset
    \label{fig:ex-cast-setup}}
\end{figure}
\begin{equation}
  \begin{split}
    \label{eq:ex-tokyo}
    &g_{a\gamma\gamma} < 6 \times 10^{-10}\; {\rm GeV}^{-1}\;\;\;{\rm
      for\;\;}m_a < 0.03\;
    {\rm eV}\;,\\
    &g_{a\gamma\gamma} < 6.8\text{--}10.9\times 10^{-10}\; {\rm
      GeV}^{-1}\;\;\;{\rm for\;\;}m_a <0.3\; {\rm eV}\;.
  \end{split}
\end{equation}
\index{Helioscope search!CAST|(}The so far most sensitive implementation of
an axion helioscope is operating at present at CERN, the CERN Axion Solar
Telescope (CAST). It was built in 2002 by refurbishing a $9.26\,\text{m}$
long prototype of a twin aperture LHC dipole magnet. The two parallel pipes
of the magnet, with a bore area of $A=14.5\,{\rm cm}^2$ each, are
homogeneously interspersed by a transverse dipole magnetic field (see Fig.
and \ref{fig:ex-cast-setup} \cite{ex-Zioutas:2004hi}). The maximum magnetic
field that can be achieved with this prototype magnet is $\approx
9\,\text{T}$. The magnet is mounted on an azimuthal moving platform that
allows to track the Sun for $1.5\,\text{h}$ ($\pm 8^{\circ}$ in height and
$\approx 100^{\circ}$ in azimuth) during sunrise and sunset.  To detect the
axions, three different types of detectors are attached to the ends of the
magnet pipes: a gaseous detector with a novel
MICROMEGAS\abbrev{MICROMEGAS}{Micromesh Gaseous Structure} readout
\cite{ex-abbon:07} and an X-ray telescope with a pn-CCD\abbrev{CCD}{Charge
  Coupled device} chip as focal plane detector \cite{ex-kuster:07} to
detect axions during sunrise, and a Time Projection Chamber
(TPC)\abbrev{TPC}{Time Projection Chamber} which covers the opposite end of
the magnet pipes, ready to collect axions during sunset
\cite{ex-autiero:07}.

CAST has taken data with the conversion region being evacuated during the
years 2003 and 2004. In the absence of any axion signal, an upper limit on
the axion to photon coupling constant has been established
to~\cite{ex-Andriamonje:2007ew,ex-zioutas:2005a}
\begin{equation}
  \label{eq:ex-cast}
  g_{a\gamma\gamma} < 0.88 \times 10^{-10}\; {\rm GeV}^{-1}\;\;\;{\rm
    for\;\;}m_a < 0.02\;{\rm eV}\;,
\end{equation}
at the 95\% confidence level. This result is the best laboratory limit from
a helioscope achieved so far. It covers a broad range of masses and
supersedes the so far best limit from astrophysical considerations that was
derived from the evolution of horizontal-branch stars in globular clusters
\cite{ex-Raffelt:2006cw,ex-Raf96}.  During 2005 the configuration of the
CAST experiment was upgraded, ready to jump into the second phase of the
experiment, when the magnet pipes are filled with $^4{\rm He}$ or $^3{\rm
  He}$ gas. In this configuration CAST will be able to expand its
sensitivity to a maximum axion mass of $\approx 1.1\,\text{eV}$. CAST is
scheduled to take data in this configuration during 2006 and 2007.
\index{Helioscope search|)} \index{Axion!Limits|)} \index{Helioscope
  search!CAST|)}

\subsection{Bragg Diffraction}
\index{Bragg diffraction|(} Another interesting and completely different
approach to detect solar axions was proposed by E.~A. Paschos and
K.~Zioutas \cite{ex-Paschos:1993yf}. This detection principle uses the
intense Coulomb field of nuclei in a crystal lattice instead of an external
magnetic field to convert axions to photons by the Primakoff effect. A
crystal can either be used simultaneously as an axion converter and
detector or to build a ``reflective parabolic antenna'' covered with a thin
film of crystals which focus the resulting photons from axion to photon
conversion on a detector in their focal plane. Background optimized solid
state detectors (e.g., NaI or Ge) generally used for WIMP searches or in
double beta decay experiments provide excellent performance to implement
the first type of a crystalline axion detector.

To calculate the probability for axion conversion, the mass of the axion
can be considered as negligibly small compared to the mass of the nuclei in
the crystal, thus recoils of the nuclei can be neglected during the
conversion process. In this case the energy of the outgoing photon would be
equal the energy of the incident axion. The differential cross section for
this conversion is given by \cite{ex-Paschos:1993yf}\index{Bragg
  diffraction!Cross section}
\begin{equation}
  \frac{\D\sigma}{\D\Omega}=\frac{g_{a\gamma\gamma}}{16\pi^2}\,F_a^2(\vec{q})\sin^2(2\Theta)\;. 
\end{equation}
Here $2\Theta$ is the scattering angle and $F_a$ a form factor which
describes the atomic structure of the crystal. The form factor $F_a$ can be
calculated from the Fourier transformation of the electrostatic field
$\Phi(\vec{x})$ in the crystal \cite{ex-cebrian:99a}
\begin{equation}
  F_a(\vec{q})=k^2 \int\Phi(\vec{x})\, e^{\imag \vec{q}\cdot\vec{x}}\,\D^3x\;,
\end{equation}
with the transfered momentum $q=|\vec{q}|=2k\sin(\Theta)$ and the axion
momentum $k=|\vec{k}|\approx E_a$. Using the Sun as an axion source, the
expected mean energy of the solar axions would be of the order of $\approx
4\,\text{keV}$ (see Fig. \ref{fig:ex-axionflux}). Thus the wave length of
the axion $\lambda$ is similar to the lattice spacing $d$ of the crystal
($\approx\text{few\,\AA}$), consequently we would expect a Bragg-reflection
pattern. If a Bragg pattern arises, the intensity of the X-rays leaving the
crystal has a maximum if the Bragg-condition\index{Bragg diffraction!Bragg
  condition}\index{Bragg condition|\\see Bragg diffraction}
\begin{equation}
  2d\sin({\Theta_\text{Bragg}})\left(1-\frac{1-n}{\sin^2(\Theta_\text{Bragg})} \right) =
  m\lambda\;\;\;\text{with}\;\;\; m = 1, 2, 3, \ldots\;,
  \label{eq:ex-bragg-condition}
\end{equation}
\begin{figure}[t]
  \centering
  \includegraphics[width=0.49\textwidth]{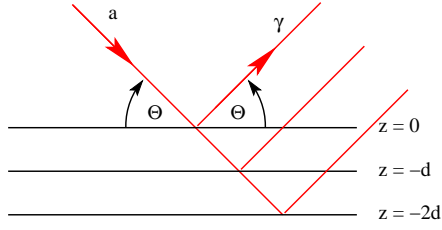}
  \caption{Schematic view of different lattice layers in a crystal. The
    axion wave enters the crystal under the angle $\Theta$. If the Bragg
    condition is fulfilled the outgoing photon waves reflected from
    different lattice layers would interfere constructively and enhance the
    expected signal\label{fig:ex-bragg-principle-principle}}
\end{figure}
is fulfilled. Here we use $n$ for the refractive index\index{Refractive
  index} and $d$ for the lattice spacing (compare Fig.
\ref{fig:ex-bragg-principle-principle}). From \eqref{eq:ex-bragg-condition}
it is obvious that constructive interference will occur only if the waves
reflected by different lattice layers are in phase. This is equivalent to
the requirement that the path length of each photon wave leaving the
crystal has to be equal to an integer multiple of its wavelength.  The
constructive interference then leads to an enhancement of the expected
signal of the order of $10^4$ compared to a scattering off a single atom in
the crystal~\cite{ex-Paschos:1993yf}.

For a detector with a fixed orientation of its crystalline lattice relative
to the Earth's surface the angle $\Theta$, which is basically the angle
between the line of sight towards the Sun and the crystalline lattice,
would change during the course of the day when the Sun moves across the
sky. Hence, the expected photon signal would have a well defined temporal
signature since the Bragg condition is only satisfied for specific times
during the day. This signature makes a potential signal clearly
distinguishable from the detector background.

\index{Bragg diffraction!Sensitivity|(} It was shown by
\cite{ex-cebrian:99a} that the maximum achievable sensitivity for an upper
limit on $g_{a\gamma\gamma}$ with this kind of detection technique is of
the order of $g_{a\gamma\gamma}\lesssim 1\times 10^{-9}\,\text{GeV}^{-1}$
using a $100\,\text{kg}$ DAMA like NaI detector \cite{ex-bernabei-99a},
assuming an observation time of $2\,\text{years}$, and a detector
background of
$2\,\text{counts}\,\text{kg}^{-1}\,\text{day}^{-1}\,\text{keV}^{-1}$. It is
obvious from this estimate, that axion detection with crystals based on
Bragg reflection is not competitive to, e.g., axion helioscopes or
astrophysical observations in terms of sensitivity with respect to
$g_{a\gamma\gamma}$. An alternative and more effective detector
configuration would be a large surface parabolic reflector (``mirror'') or
an array of reflectors covered with a thin crystalline
layer~\cite{ex-Paschos:1993yf}. In addition polar crystals (ferroelectric)
materials can improve the sensitivity, since they provide stronger
microscopic electric fields \cite{ex-Paschos:1993yf}. Nevertheless, the
Bragg reflection technique has a major advantage, its sensitivity does not
dependent on the mass of the axion, as long as nuclear recoils can be
neglected.  \index{Bragg diffraction!Sensitivity|)}

\index{Axion!Limits|(}
Two different collaborations, COSME \cite{ex-Morales:2001we} and SOLAX
\cite{ex-Avignone:1997th}, both using Germanium detectors with the main
purpose of searching for Dark Matter WIMPs, analyzed their data looking for
an axion signature. Both experiments were able to yield very similar
mass-independent bounds to the axion to photon coupling in the absence of
any signal\index{SOLAX}\index{COSME}
\begin{equation}
  \begin{split}
    \label{eq:ex-cosme}
    &g_{a\gamma\gamma} < 2.7 \times 10^{-9}\; {\rm GeV}^{-1}\;\;\;({\rm SOLAX})\;,\\
    &g_{a\gamma\gamma} < 2.8 \times 10^{-9}\; {\rm GeV}^{-1}\;\;\;({\rm COSME})\;.\\
  \end{split}
\end{equation}
The DAMA collaboration~\cite{ex-Bernabei:2001ny} achieved a similar limit\index{DAMA}
\begin{equation}
  \label{eq:ex-dama}
  g_{a\gamma\gamma} < 1.7 \times 10^{-9}\; {\rm GeV}^{-1}\;\;\;({\rm DAMA})\;,
\end{equation}
using NaI(Tl) crystals as detector material.
\begin{figure}[t]
  \centering
  \includegraphics[width=0.78\textwidth]{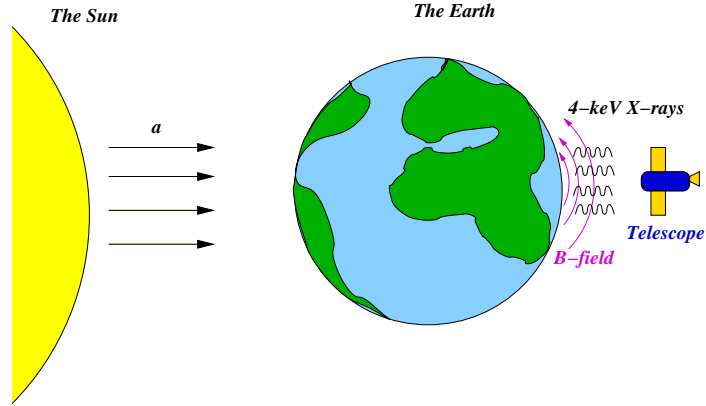}
  \caption{\label{fig:ex-gecosax} The observational setup to detect
    geomagnetic converted axions (GECOSAX). The Sun is used as an axion
    source and a satellite based X-ray observatory to detect the photons
    from geomagnetically converted axions. The Earth acts as a shield for
    direct X-rays coming from the Sun and other X-ray sources on the sky}
\end{figure}
\index{Axion!Limits|)}
\index{Bragg diffraction|)}

\subsection{Geomagnetic Axion Conversion}
\index{Geomagnetic axion conversion|(} It has recently been shown by
\cite{ex-Davoudiasl:2005nh} that Geomagnetic Conversion of Solar Axions to
X-rays (GECOSAX)\abbrev{GECOSAX}{Geomagnetic Conversion of Solar Axions}
can yield a photon flux which is measurable by a satellite based X-ray
observatory on the dark side of the Earth. A key ingredient of this idea is
to use the Earth as an X-ray shield and look for axions on its dark side.
Using the Earth as a ``shield'' towards the Sun, effectively removes the
solar X-ray background (see Fig.  \ref{fig:ex-gecosax} for a schematic
representation of this setup). In the following section, the feasibility of
such an experiment will be demonstrated, based on recent measurements of
the dark-Earth X-ray background by the Suzaku satellite.

The radius of the Earth is $\re \approx 6.4\times10^3\,\text{km}$ and its
magnetic field is well approximated by a dipole for distances less than
$1000\,\text{km}$ above the surface. The magnetic field strength at the
equator is $\be \simeq 3\times10^{-5}\,\text{T}$ and it drops as $\propto
1/r^3$~\cite{ex-Landolt}.  However, over distances $L \ll \re$ above the
Earth's surface, it can be assumed that $\be = \text{const}$.  As we are
interested in low Earth orbits ($L < 1000\,\text{km}$), this is a valid
assumption \cite{ex-Davoudiasl:2005nh}. The influence of the atmosphere of
the Earth is negligible, since for $L > 150\,\text{km}$ solar axions
essentially travel in vacuum.  For an axion mass
$m_a\leq10^{-4}\,\text{eV}$, a mean axion energy $E_a = 4\,\text{keV}$, and
a satellite orbit of $L_\oplus \simeq 600\,\mathrm{km}$, the axion to
photon conversion probability can be approximated as\index{Geomagnetic
  axion conversion!Conversion probability}
\begin{equation}
   P_{a\to\gamma}(L) = \frac{1}{4}\left(g_{a \gamma\gamma}B L\right)^2\;,
  \label{eq:ex-gecosax-probability}
\end{equation}
where $g_{a \gamma\gamma}$ is the coupling strength of the axion to the
photon.
\begin{figure}[t]
  \centerline{\includegraphics[width=0.8\columnwidth]{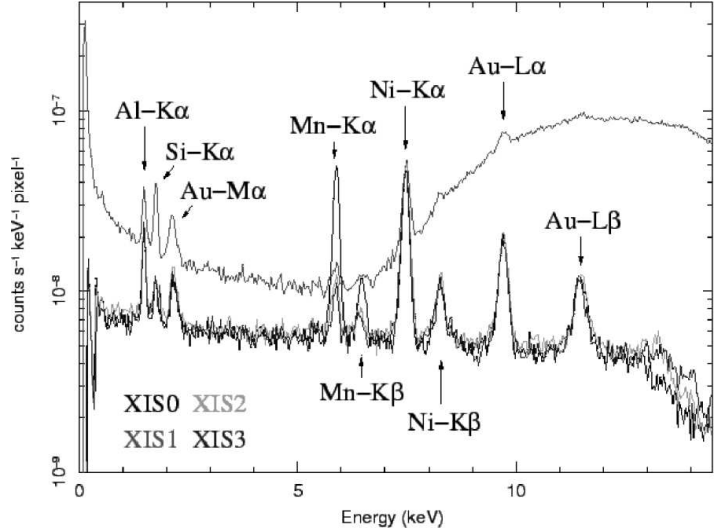}}
  \caption{\label{fig:ex-dark} 
    Dark-Earth background measured by Suzaku based on a $800\,\text{ksec}$
    long observation. Individual fluorescent emission lines originating in
    detector materials are marked \cite{ex-suzakufig}}
\end{figure}
\begin{figure}[t]
  \centering
  \includegraphics[width=0.75\textwidth]{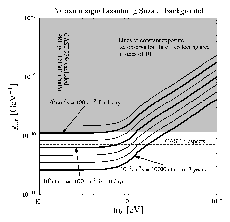}
  \caption{\label{fig:reach}
    The shaded area schematically depicts the CAST $95\%$ C.L. excluded
    region~\cite{ex-Andriamonje:2007ew}. The lines show sensitivities
    obtainable with the background measured by Suzaku}
\end{figure}
To be able to compare the expected axion flux in a specific energy range
with available data, the spectrum given by
\eqref{eq:ex-solar-axion-spectrum} is integrated over the energy range of
$2$ to $7\,\text{keV}$ to find the solar axion flux at Earth to
be\index{Axion!Solar!Spectrum}
\begin{equation}
  \Phi_a \approx 2.7 \times 10^{11} (g_{a\gamma}/10^{-10}\, {\rm
    GeV}^{-1})^2\,{\rm cm}^{-2}\,{\rm s}^{-1}\;.
  \label{eq:ex-Phia}
\end{equation}
Thus, using \eqref{eq:ex-gecosax-probability} the expected flux of X-rays
from axion to photon conversion at an altitude of $L_\oplus$ near the
equator for $g_{a \gamma} = 10^{-10}\,\text{GeV}^{-1}$ and $B=\be$ is
\begin{equation}
  \Phi_\gamma(L_\oplus) \equiv \Phi_a \,P_{a\to\gamma}(L_\oplus) \approx 3\times
  10^{-7}\,{\rm cm}^{-2}\,{\rm s}^{-1}\;. 
  \label{eq:ex-Phigam}
\end{equation}
\index{RXTE}Based on the published technical specifications \cite{ex-rxte}
of the Rossi X-ray timing explorer (RXTE)\abbrev{RXTE}{Rossi X-ray timing
  explorer}, \cite{ex-Davoudiasl:2005nh} have estimated that the flux given
by \eqref{eq:ex-Phigam} can be detected on the dark side of the Earth.

\index{Suzaku} However, recently the Suzaku X-ray satellite team has
measured the dark-Earth X-ray background in the range of $2$ to
$7\,\text{keV}$ for calibration purposes \cite{ex-suzaku}. Suzaku has a
circular orbit at $575\,\text{km}$ with an inclination of $33^\circ$.
Unfortunately, the Suzaku satellite does not provide a fast slew rate and
thus cannot perform this kind of measurement.  However, as will be shown
below, a telescope, capable of tracking the solar core on the dark side of
the Earth and with similar X-ray detection capabilities will be sensitive
to axion-photon couplings $g_{a\gamma\gamma}$ far below the current
existing laboratory and astrophysical bounds for axion masses $m_a <
10^{-4}\,\text{eV}$.

The expected signal from axion to photon conversion has several distinct
features which will allow to clearly distinguish it from the detector
background. These features are
\begin{enumerate}
\item a thermal like spectral distribution of X-rays peaked at
  approximately $4\,\text{keV}$, on the night side of the Earth,
\item the X-rays would only come from the center of the Sun, which subtends
  approximately $3'$ on the sky,
\item the observed X-ray intensity would vary with the magnetic field
  strength and thus with the orbital position of the satellite and its
  orientation relative to the Earth's magnetic field (pointing direction of
  the satellite), and
\item the signal would be modulated with the Sun-Earth distance, resulting
  in an annual modulation of the intensity of the signal.
\end{enumerate}
\index{Geomagnetic axion conversion!Sensitivity} To estimate the
sensitivity of Suzaku for solar axion detection, the effective area of the
Suzaku detector was assumed to be $\approx 300\,\text{cm}^2$ in the
$2$--$7\,\text{keV}$ range~\cite{ex-suzaku}. The expected dark Earth
background of the Suzaku XIS\abbrev{XIS}{X-ray Imaging Spectrometer}
detector per unit area is presented in Fig.  \ref{fig:ex-dark}
\cite{ex-suzakufig}.  Given the above effective area, the $\approx 1'$
resolution of the X-ray telescope and the Suzaku dark Earth background
data, the measurement of the X-ray flux from solar axion conversion is
clearly feasible.  The Suzaku team estimates that with $3\times
10^5\,\text{s}$ of data, a $4\sigma$ bound $g_{a \gamma\gamma} <
10^{-10}\,\text{GeV}^{-1}$ would have been possible \cite{ex-suzaku}. Based
on the Suzaku background data \cite{ex-suzaku}, the sensitivity of a
solar-core-tracking X-ray telescope is plotted as a function of $m_a$ in
Fig. \ref{fig:reach}. The Earth-occulted background can in general be
measured in situ, by pointing the X-ray telescope away from the direction
of the core of the Sun. The observation of such a signal amounts to viewing
an X-ray image of the solar core through the Earth. Therefore, this method
can achieve an unambiguous detection of solar axions. In summary, for solar
axions with $m_a < 10^{-4}\,\text{eV}$, an orbiting X-ray telescope, with
an effective area of a few $10^3\,\text{cm}^2$, can probe solar axion $g_{a
  \gamma\gamma}$ well beyond the sensitivity of current laboratory
experiments.  \index{Geomagnetic axion conversion|)}

\section{Searches for Laser Induced Axions}
\index{Laser induced axions|(} In this section, a new generation of purely
laboratory based experiments which are able to provide complementary
results to solar axions searches, will be briefly introduced . Whereas the
solar axion experiments, described in the previous sections, probe axions
which would escape from the Sun, the experiments presented in the following
are supposed to produce axions from polarized laser beams, propagating in a
transverse magnetic field. Yet the mechanism for both, the production and
detection, is the same as in the previously described experiments: the
Primakoff effect.

Searches for laser induced axions provide a fully model independent
approach, since they do not rely on the physical processes and conditions
in the Sun and how axions or axion-like particles could be produced under
these conditions. They can be viewed as fixed target type experiments, with
low energy polarized photon beams colliding with virtual photons provided
by the magnetic field, i.e., $\gamma + \gamma_\text{virtual}\rightarrow\ 
a$. By pushing the optical detection techniques up to the present
state-of-art, a significant enlarged domain of axion mass and axion
di-photon coupling constant, not yet explored with laser based experiments,
is expected to be probed with photons, typically in the energy range of a
few $\text{eV}$. In addition, laser based experiments are not only focused
on the axion search, but offer also a broad band of scientific interests,
starting from a precise new test of QED up to the search of any scalar,
pseudo-scalar, or other particles like paraphotons or millicharged
particles that can couple to photons. In general, laser based experiments
can be divided into two categories,
\begin{itemize}
\item the so called ``photon regeneration'' or ``light shining through a
  wall'' experiments and
\item experiments that probe the magneto-optical properties of the vacuum.
\end{itemize}

\subsection{Light Shining Through a Wall}
\begin{figure}[t]
  \centering
  \includegraphics[width=0.8\textwidth]{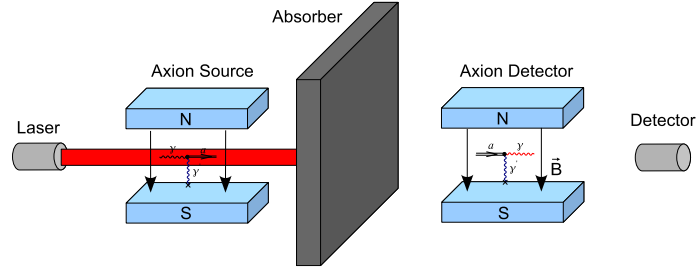}
  \caption{Schematic view of a photon regeneration or light shining
    through a wall experiment. A polarized laser beam enters a transverse
    magnetic field from the left. A small fraction of the laser photons can
    be transmuted to, e.g., axions.  While the laser light is blocked by a
    thick absorbing wall, the axions would pass it. A second magnetic field
    behind the wall is used to reconvert the axions to photons which are
    finally detected \cite{ex-VanBibber:1987rq}
    \label{ex-light-shining-principle}}
\end{figure}
The light shining through a wall or photon regeneration principle was first
proposed by \cite{ex-VanBibber:1987rq} in 1987 (for an independent proposal
of this type, see \cite{ex-Anselm:1986gz}).  The basic idea of this
experiment is shown in Fig.  \ref{ex-light-shining-principle}.  A polarized
laser beam propagates inside a transverse magnetic field (with {\vec{E}}
$||$ {\vec{B}}). The laser is blocked by a wall at some point of its path,
such that only weakly interacting pseudoscalar or scalar particles created
before will be able to pass through this absorber. The conversion
probability for the production of an axion-like particle in this case is
given by
\begin{equation}
    P_{\gamma\to a}\propto \frac{1}{4}(g_{a\gamma\gamma} BL)^2\,F(qL)\;,
\end{equation}
were $F(qL)$ is the form factor given by \eqref{eq:ex-probability}. A
fraction of these weakly interacting particles will turn back into
detectable photons after they have passed the wall and enter the second
magnetic field, called ``Axion Detector'' in Fig.
\ref{ex-light-shining-principle}.  The total probability to observe a
regenerated photon with a detector located at the end of the second
magnetic field is
\begin{equation}
  P_{\gamma\to a \to\gamma}= P^2_{\gamma\to a}\;.
\end{equation}
The expected counting rate depends on the power of the laser $P$ and can be
enhanced by an optical cavity installed on the production side. If the
total number of reflections in the primary magnet is $N_\text{r}$ and the
detection efficiency is $\eta$, then the total number of counts expected
from reconversion calculates from \cite{ex-ringwald:2005a}
\begin{eqnarray}
  \frac{\D N_\gamma}{\D t} &=& \frac{1}{16} \frac{\langle P
   \rangle}{\omega} \frac{N_\text{r}+2}{2}\, \eta\, (g_{a\gamma\gamma} BL)^4
   \sin^2\left(\frac{m_a^2L}{4\omega}\right)/\left(\frac{m_a^2
       L}{4\omega}\right)^4\\ 
  &\approx & \frac{1}{16} \frac{\langle P \rangle}{\omega}
   \frac{N_\text{r}+2}{2}\, \eta\, (g_{a\gamma\gamma} BL)^4
   \;,\label{eq:ex-approximation} 
\end{eqnarray}
where the approximation \eqref{eq:ex-approximation} is valid only for $m_a
\ll \sqrt{2\pi\omega/L}$.

The pioneering experiment based on this technique was performed by the
Brookhaven-Fermilab-Rutherford-Trieste (BFRT) collaboration.  They used two
superconducting dipole magnets of length $L=4.4\,\text{m}$ which were able
to provide a magnetic field of $B=3.7\,\text{T}$. The optical laser with a
wavelength of $\lambda = 514\,\text{nm}$ had an average power of $\langle
P\rangle = 3\,\text{W}$ and was operated in an optical cavity providing
$200$ reflections. Since no signal from photon regeneration was found, the
BFRT collaboration was able to set an upper limit on $g_{a\gamma\gamma}$ of
\cite{ex-Cameron:1993mr}
\begin{equation}
  \label{eq:ex-shine}
  g_{a\gamma\gamma}<6.7 \times 10^{-7}\,\text{GeV}\,\,
  (95\%\,\text{C.L.})\;,  
\end{equation}
for axion-like pseudoscalars with a maximum mass of
\begin{equation}
  \label{eq:ex-2shine}
  m_a < 10^{-3}\,\text{eV}\;.
\end{equation}

\begin{figure}[t]
  \centering
  \includegraphics[width=0.7\textwidth]{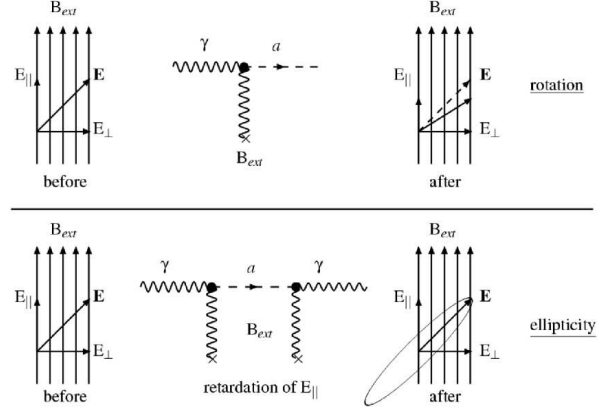}
  \caption{Up: Linear Dichroism or rotation of the polarization vector by
    and angle $\varepsilon$. Down: Linear Birefringence or induction of an
    ellipticity $\Psi$ in an initially linear polarized beam}
  \label{fig:ex-light-propierties}
\end{figure}
\subsection{Probing the Magneto-Optical Properties of the Vacuum}
The second of the purely laboratory experimental approaches to search for
axions and axion-like particles is of indirect detection type. It is based
on the theoretical prediction that scalar or pseudo-scalar particles can
affect the polarization of light propagating in vacuum through a transverse
magnetic field $\vec{B}$ because of their coupling to photons
\cite{ex-Maiani} (see also Chap. \ref{chap:cantatore-results-from-pvlas}).
A light beam initially linearly polarized with an angle $\theta$ with
respect to $\vec{B}$ is expected to acquire a small ellipticity $\Psi$ and
a small apparent rotation $\Theta$ due to dispersive and absorptive
processes induced by the production of spinless particles. Both effects are
usually referred to as birefringence and as linear dichroism of the vacuum
and are characterized by
\begin{equation}\label{ex-eq1}
  \Psi \approx N~\frac{B^2 L^3 m_{A}^2}{96 \omega M^2}\,\sin(2\theta)\;,
\end{equation}
\begin{equation}\label{ex-eq2}
  \Theta \approx N~\frac{B^2 L^2}{16~M^2}\,\sin(2\theta)\;,
\end{equation}
in the limit $m_{A}^{2}L/4\omega\ll 1$. Here $m_{A}$ is the axion mass,
$M=1/g_{a\gamma\gamma}$ the inverse coupling constant to two photons,
$\omega$ the photon energy, $L$ and $N$ the effective path lengths and the
number of paths the light travels in the transverse magnetic field. The
polarization state of the light beam leaving the magnetic field,
characterized by \eqref{ex-eq1} and \eqref{ex-eq2}, would manifest itself
as a sizeable deviation from the pure QED prediction \cite{ex-QED,ex-VMB}
for which no measurable linear dichroism is expected.

The photon splitting effect can also produce a differential absorption
\cite{ex-Adler}, giving rise to an apparent angular rotation of the
polarization of the order of $10^{-34}\,\text{rad}$ in a $9.5\,\text{T}$
magnetic field over a length of $25\,\text{km}$. This angular rotation
angle is far from being measurable under laboratory conditions, except if
the coupling with scalar or pseudo-scalar particles would significantly
enhance this effect \cite{ex-Gabrielli}.

The vacuum magnetic birefringence predicted by the QED, corresponds to the
dispersive effect produced by virtual electron-positron pairs as this was
already stated by Heisenberg and Euler in 1936 \cite{ex-QED}: ``Even if
electromagnetic fields are not strong enough to create matter, they will,
due to the virtual possibility of creating matter, polarize the vacuum and
therefore change the Maxwell's equations''. This ellipticity constitutes
the background signal for a measurement that aims to detect a birefringence
or dichroism induced by a spinless particle. The ellipticity expected from
vacuum-polarization can be expressed from one loop calculation as
\cite{ex-QED,ex-VMB}
\begin{equation}\label{eq3}
  \Psi _\text{QED}\approx N~\frac{B^2 L \alpha^2
    \omega}{15\,\text{m}^{4}}\sin(2\theta)\;, 
\end{equation}
where $\alpha \approx 1/137$ is the fine-structure constant, $\omega$ is
the photon energy and $m_e$ the electron mass. The maximum of ellipticity,
a laser beam with a wavelength $\lambda = 1550\,\text{nm}$ propagating in a
$9.5\,\text{T}$ field over a length $NL = 25\,\text{km}$ would acquire is
equal to $2\times10^{-11}\,\text{rad}$.

A search for both effects, birefringence and dichroism, was carried out
with the same magnets used in the BFRT experiment \cite{ex-Cameron:1993mr},
setting a bound on the axion to photon coupling constant of
\begin{equation}
  \label{eq:ex-pol}
  g_{a\gamma\gamma} <3.6 \times 10^{-7}\, {\rm GeV}^{-1}\;,
\end{equation}
for masses 
\begin{equation}
  \label{eq:ex-2pol}
  m_a <5\times 10^{-4}\,{\rm eV}\;,
\end{equation}
at the 95\% C.L. 

Very recently the Italian PVLAS experiment \cite{ex-Zavattini:2005tm} (see
Chap. \ref{chap:cantatore-results-from-pvlas}) has been taking data to test
the vacuum birefringence in the presence of a magnetic field with a
$1\,\text{m}$ long dipole magnet operated at a maximum field of
$5.5\,\text{T}$, therefore improving the sensitivity of the previous
experimental setups.  For the first time the PVLAS collaboration has
measured a positive value for the amplitude of the rotation $\varepsilon$
of the polarization plane in vacuum with $B \approx 5\,\text{T}$ (quoted
with a $3\sigma$ uncertainty interval) \cite{ex-Zavattini:2005tm}
\begin{equation}
  \label{eq:ex-pvlas-rot}
  \varepsilon=(3.9\pm 0.5)\times 10^{-12}\,\text{rad}\,\text{pass}^{-1}\;.
\end{equation}
This signal can be translated to an allowed range for the mass $m_b$ and
the coupling constant to two photons $g_{b\gamma\gamma}$ of a neutral light
pseudoscalar boson
\begin{eqnarray}
  \label{eq:ex-boson}
  1 \,\mbox{meV} \le &m_b&  \le 1.5 \,\mbox{meV}\nonumber\;,\\
  1.7 \times 10^{-6} \,\mbox{GeV}^{-1} \le  &g_{b\gamma\gamma}& \le 1
  \times 10^{-5}\,\mbox{GeV}^{-1}\;. 
\end{eqnarray}
Several new experiments are planned and already in progress with the goal
to verify this signal. They are based on both experimental techniques: the
light shining through a wall principle and the principle to test the
properties of the quantum vacuum or a combination of both. In Table
\ref{tab:ex-wall}, a summary of actually planned or already operating
experiments is given together with their performance parameters. A more
detailed introduction to some of the experiments quoted in Table
\ref{tab:ex-wall} is given in the following sections.
\begin{table}[t]
\caption{Planned laser based experiments to detect
    axion-like particles. The most important magnet and laser parameters are
    shown. PVLAS, and later also OSQAR, will use optical cavities to
    enhance the number of generated  ALPs. $P_{\gamma\Phi\gamma}$ (PVLAS)
    denotes the probability for a photon-ALP-photon conversion for the
    different experimental setups based on the coupling constant derived
    from the PVLAS measurement and the BFRT limits. The last column gives
    the expected signal rate of re-converted photons \label{tab:ex-wall}} 
  \begin{center}
    \begin{tabular}{p{1.5cm}p{1.8cm}p{2cm}p{2cm}rr}\hline
Name    & Location      &\multicolumn{1}{c}{Laser}      &\multicolumn{1}{c}{Magnet}     &\multicolumn{1}{c}{$P_{\gamma\phi\gamma}$} & \multicolumn{1}{c}{$\gamma$ Flux} \\
        &               &                               &                               &(PVLAS)                & \\[0.5ex] \hline\hline
\rule[0mm]{0mm}{4mm}ALPS    &      DESY/D   & $\lambda=1064\,\text{nm}$     &$B = 5 \,\text{T}$             &$\sim 10^{-19}$        & $10^1\,\text{s}^{-1}$\\ 
        &               & $P = 200\,\text{W}$           &$L = 4.21\,\text{m}$           &                       & \\ 
BMV     &       LULI/F  & $\lambda = 1053\,\text{nm}$   &$B = 11\, \text{T}$            &$\sim 10^{-21}$        & $10\,\text{pulse}^{-1}$\\ 
        &               & $P = 500\,\text{W}$           &$L = 0.25\,\text{m}$         &                       & \\ 
LIPSS   &  Jlab/USA     & $\lambda=900\,\text{nm}$      &$B = 1.7 \,\text{T}$           &$\sim 10^{-23.5}$      & $10^{-1}\,\text{s}^{-1}$\\ 
        &               & $P = 10\,\text{kW}$           &$L = 1\,\text{m}$              &                       & \\ 
OSQAR   &   CERN/CH     & $\lambda= 540\,\text{nm}$     &$B = 9.5\,\text{T}$            &$\sim 10^{-18}$        & $10^{3}\,\text{s}^{-1}$\\
        &               & $P =1\,\text{kW}$             &$L_1 = 1 \,\text{m}$           &                       & \\ 
        &               &                               &$L_2= 13.3\,\text{m}$          &                       & \\ 
PVLAS   &  Legnaro/I    & $\lambda=1064\,\text{nm}$     &$B_1=5\,\text{T}$              &$\sim 10^{-23}$        & $10^{-1}\,\text{s}^{-1}$\\
        &               & $P=0.8\,\text{W}$             &$B_2=2.2\,\text{T}$            &                       & \\ 
        &               &                               &$L_1=4.21 \,\text{m}$          &                       & \\
        &               &                               &$L_2=0.5 \,\text{m}$           &                       & \\ \hline
\end{tabular}
\end{center}
\end{table}

\subsubsection{Axion-Like Particle Search -- ALPS}
\begin{figure}[t]
  \centering
  \includegraphics[width=0.9\textwidth]{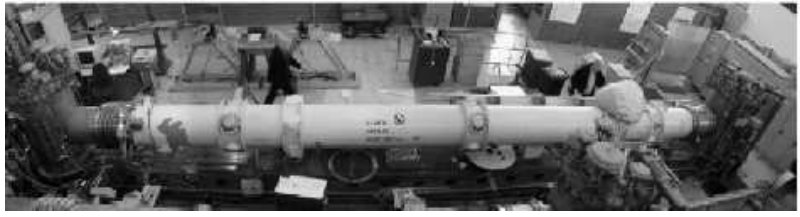}
  \caption{Superconducting HERA dipole magnet exploited by ALPS for conversion 
    of laser photons into axion-like particles, as well as for the
    reconversion of axion-like particles into photons}
  \label{fig:ex-heradipole}
\end{figure}
A collaboration of DESY, Laser Zentrum Hannover and Sternwarte Bergedorf is
presently setting up the Axion-Like Particle Search (ALPS) photon
regeneration experiment which exploits the photon beam of high-power
optical lasers, sent along the transverse magnetic field of a
superconducting HERA dipole magnet \cite{ex-Ehret:2007cm}. This experiment,
which has been approved by the DESY directorate on January 11, 2007, and is
expected to take data in summer 2007, offers a window of opportunity for a
rapid firm establishment or exclusion of the axion-like particle
interpretation of the anomaly published by PVLAS. In case of confirmation,
it would also allow for the measurement of mass, parity, and coupling
strength of this particle.

\begin{figure}
  \centering
  \includegraphics[width=0.9\textwidth]{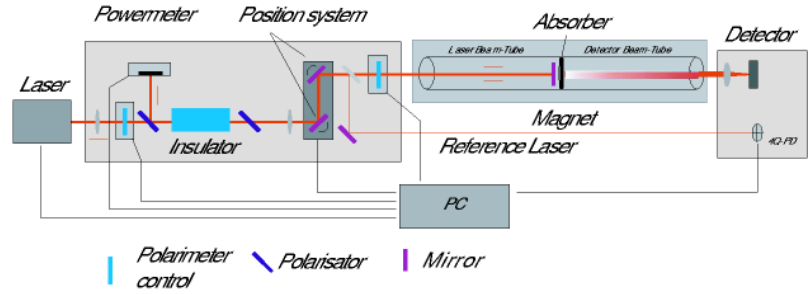}
  \caption{\label{fig:ex-layout}
    Schematic view of the experimental setup with the laser on the left,
    followed by the laser injector/extractor system, the magnet and the
    detector table. An intensity-reduced reference beam of the laser is
    guided parallel to the magnet for constant alignment monitoring between
    laser and detector}
\end{figure}

The photon regeneration experiment ALPS is build around a spare dipole of
the HERA proton storage ring at the DESY magnet test stand (see Fig.
\ref{fig:ex-heradipole}).  Both parts of the experiment, i.e., axion-like
particle production and reconversion into photons are accommodated in one
single magnet, since the test stand architecture in its present
configuration forbids to place two fully functional magnets in line. The
general layout of the experiment is depicted in Fig.  \ref{fig:ex-layout}.
A high intensity laser beam is placed on one side of the magnet traversing
half of its length. In the middle of the magnet, the laser beam is
reflected back to its entering side, and an optical barrier prevents any
photons from reaching the second half of the magnet.  Axion-like particles
would penetrate the barrier, eventually reconverting into photons inside
the second half of the magnet. Reconverted photons are then detected with a
pixeled semi-conductor detector outside the magnet.

The magnetic field will be produced by a spare dipole magnet of the HERA
proton storage ring (see Fig. \ref{fig:ex-heradipole}).  At a nominal
current of $6000\,\text{A}$, the magnet reaches a field of $5.4\,\text{T}$
over a total magnetic length of $8.82\,\text{m}$.  In a first stage, ALPS
will exploit a laser system delivering a linearly polarized photon beam
with an average power of $50\,\text{W}$ at a wavelength of $532\,\text{nm}$
and a low noise, high quantum efficiency commercial CCD camera.  Already
with this configuration, the axion-like particle interpretation of PVLAS
can be clarified: in this case, the expected counting rate is about
$2\,\text{Hz}$.  In a second stage, it is planned to use a commercial fiber
laser (Nd:YAG) at $1064\,\text{nm}$ which is able to deliver
$200\,\text{W}$ linearly polarized photons, together with a InGaAs pixel
detector normally used in infrared astronomy. This configuration will
enable ALPS to explore also ``new territory'' in the coupling vs. mass
plot.

\subsubsection{The BMV Project}
The BMV (Birefringence Magnetique du Vide) project \cite{ex-BMV}, which is
being built in Toulouse, France, combines very intense pulsed magnetic
fields, developed at LNCMP, and a very sensitive optical device to detect
the effects induced on a laser beam by such fields. This device is
developed at LCAR-IRSAMC in collaboration with LMA-VIRGO from IN2P3 in
Lyon, France. The goal was to assemble a first version of the experiment in
2006, and to obtain first results in 2007. In particular, the setup will be
able to test the results published by the PVLAS
collaboration~\cite{ex-Zavattini:2005tm}.

The BMV project is based on a Fabry-P\'erot cavity with a finesse of about
$1\times 10^6$ and a pulsed magnetic field as high as $25\,\text{T}$ over
the length of about one meter. The sensitivity limit for the measured
ellipticity is about $4\times 10^{-9}$, more than 100 times the one
expected with the PVLAS apparatus. In the final stage, the experiment will
have the sensitivity to measure the vacuum birefringence as predicted by
QED.

\begin{figure}[t]
  \begin{center}
    \includegraphics[width=0.95\textwidth]{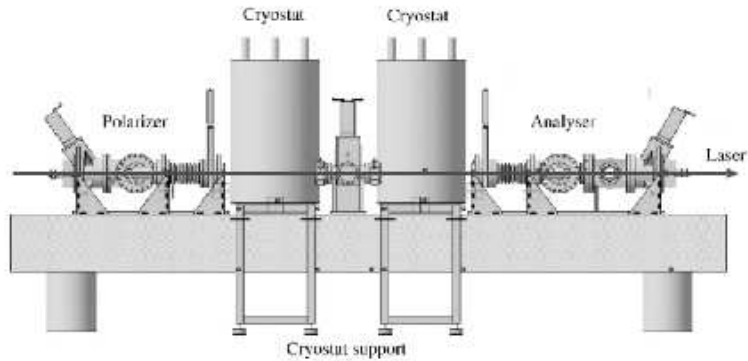}
    \vspace{-1.2cm}
    \caption{Schematic view of the experimental set up of the BMV
    project\label{fig:ex-exp}} 
  \end{center}
\end{figure}

The experimental setup of the optical apparatus is shown in Fig.
\ref{fig:ex-exp}. The light provided by a Nd:YAG laser ( $\lambda
=1064\,\text{nm}$) laser is polarized and it is injected in a
Fabry-P\'{e}rot cavity containing the magnetic field region. The light
exiting the cavity is analyzed by a polarizer and detected by a photodiode.
Light reflected back by the cavity is collected and this signal is used to
drive the locking electronics that changes the laser frequency to keep the
Fabry-P\'{e}rot resonating. The measurement will be performed by phase
detection, since the magnetic field is pulsed.
\begin{figure}[t]
  \begin{center}
    \includegraphics[width=0.75\textwidth]{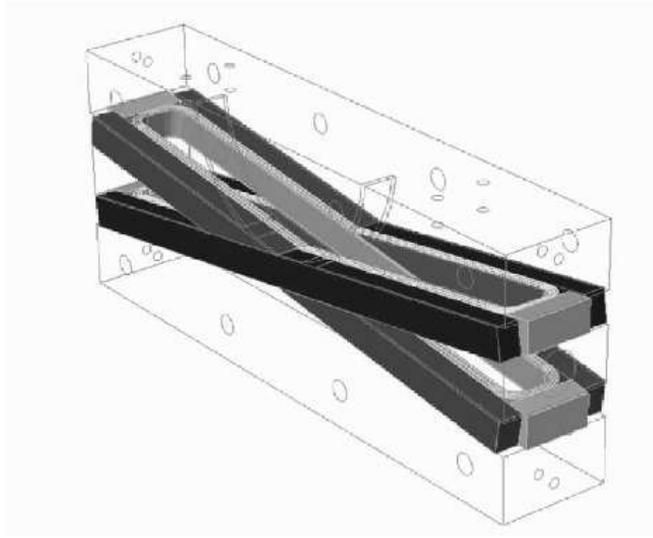}
    \caption{Schematic view of the X-coil used in BMV\label{fig:ex-proto}} 
  \end{center}
\end{figure}

\begin{figure}[t]
  \begin{center}
    \includegraphics[width=0.90\textwidth]{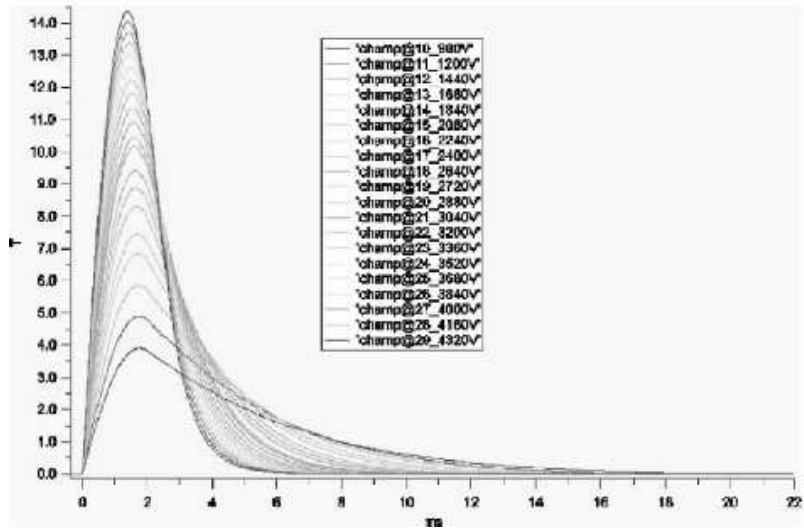}
    \caption{Magnetic field delivered by the 
      X-coil depending on time and for different supply voltages
      \label{fig:ex-B}}
  \end{center}
\end{figure}

To achieve a maximum in sensitivity, the project demands a transverse
magnetic field as high as possible. This can only be obtained using pulsed
magnet techniques. Moreover, a magnetic field region as long as possible is
required, since the QED effect depends on the product $B^2L$, where $L$ is
the length of the magnetic field region. The goal is to realize a pulsed
magnet delivering a transverse magnetic field approaching $25\,\text{T}$.
Due to the high field value the structure of the magnet is very constraint.
In particular, at this field level, the magnetic pressure corresponds to
$250\,\text{MPa}$ ($2.5\,\text{t}\,\text{cm}^{-2}$). Pulsed magnets that
can provide such high fields have already been developed and tested, based
on the X-coil geometry shown in Fig. \ref{fig:ex-proto}.  The length of the
coil is $0.25\,\text{m}$, providing a peak field of $14.3\,\text{T}$
corresponding to about $28\,\text{T}^2\,\text{m}$ (see Fig.
\ref{fig:ex-B}).
This is one of the highest field strengths ever reached with such a
configuration. One of the prototype coils has been aged at a duty cycle of
about $5$ pulses per hour. It has delivered $100$ pulses at a peak field of
$11.5\,\text{T}$ and $100$ pulses at a peak field of $12.5\,\text{T}$,
corresponding to about $21\,\text{T}^2\,\text{m}$. The magnet will be
supplied by a pulsed generator based on a bank of 12 capacitors. In
particular, this generator will be able to produce a single pulse or to
give a field that oscillates.

As far as the optical system is concerned, a system of high precision
mechanical translators and rotators for the Fabry-P\'{e}rot cavity mirrors
and for the polarizers has been designed and assembled at LCAR in Toulouse.
The piezoelectric stacks of the mirrors orientation system, as well as the
mechanism allowing their rotation are adapted to ultra-high vacuum. Tests
have been carried out with a vertical $3.6\,\text{m}$ long Fabry-P\'{e}rot
cavity based on the LMA mirrors, made to reach a finesse around
$5\times10^{5}$.
Figure \ref{fig:ex-exp} shows a schematic drawing of the experiment with
the Fabry-P\'{e}rot resonator, the magnetic field region, and two
cryostat. The chambers for the optics operated at a ultra high vacuum, are
fixed on an optical table with a length of $3.6\,\text{m}$. The length of
the cavity is about $2\,\text{m}$.

2006 was a crucial year for the project. In May that year the experiment
will start to be set up at LNCMP. First with just one magnet in place,
then, when test runs will be completed, with two magnets in place. This
configuration corresponds to $40\,\text{T}^2\,\text{m}$. A new set of
mirrors is in preparation at LMA.  The clean environments, already used at
LCAR, moved in the experimental clean room at LNCMP will allow to exploit
at their best these new mirrors. Finesse greater than 200 000 is expected.
The sensitivity should also be at least $10^{-8}\times$ 1/$\sqrt(Hz)$,
thanks to the high central frequency of the modulated effect.

\begin{figure}[t]
  \begin{center}
    \label{fig:ex-axion2}
    \includegraphics[width=0.8\textwidth]{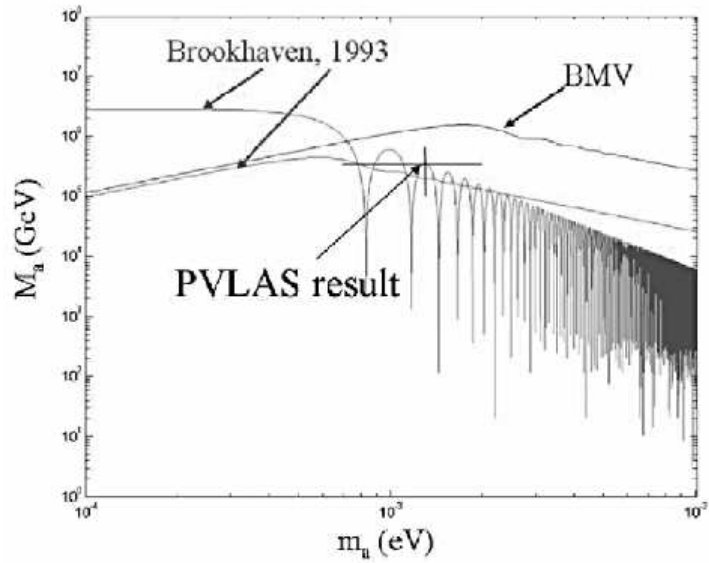}
    \caption{Limits on the inverse coupling constant
      $M_a$ in the $m_a$ vs. $M_a$ plane for an axion-like particle given
      for one day of operation (20 magnet pulses) assuming a zero
      ellipticity measurements}
  \end{center}
\end{figure}

Fig. \ref{fig:ex-axion2} represents the exclusion region in the $m_a$ vs.
the inverse coupling constant $M_a$ plane, given for one day of operation
(20 magnet pulses) and assuming a zero ellipticity measurement, based on
the experimental parameters given above. The red curves indicate the
Brookhaven (BFRT) result \cite{ex-Cameron:1993mr}, the blue line the BMV
projected result, and finally the black cross the region of values that
could explain the PVLAS results. This interpretation would thus be excluded
by a zero ellipticity measurement with the BMV setup.

\subsubsection{The OSQAR experiment at CERN (2-in-1 Experiment)}
\begin{figure}[t]
  \centering
  \includegraphics[width=0.7\textwidth]{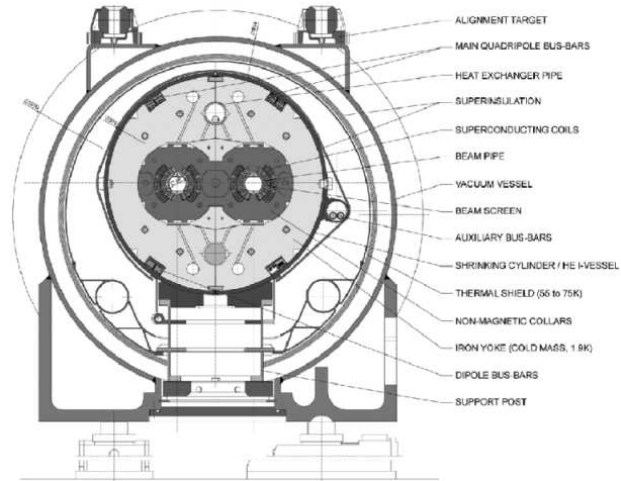}
  \caption{Cross section of a LHC superconducting main dipole magnet
    housed inside its cryostat} \label{layout}
\end{figure}

Because of the strong transverse magnetic field, required to obtain
measurable effects, an ideal implementation to investigate simultaneously
the magneto-optical properties of the quantum vacuum and the photon
regeneration effect, are within long superconducting accelerator dipolar
magnets, such as the ones developed and manufactured for the Large Hadron
Collider (LHC). The re-use of recently decommissioned $15\,\text{m}$ long
twin aperture LHC superconducting magnet prototypes, providing a transverse
magnetic field $B=9.5\,\text{T}$, offers a unique opportunity for the
construction of a new powerful two-in-one experiment, based on optical
techniques \cite{ex-Pugnat}. The cross section of such dipole magnet is
shown in Fig. \ref{layout}.

Linearly polarized laser light beams will be used as probes inside vacuum
chambers housed inside superconducting magnet apertures. One of the
apertures will be dedicated to the measurements of the Vacuum Magnetic
Birefringence (VMB) and optical absorption anisotropy whereas the other one
will be used to detect the photon regeneration from axions or axion-like
particles using an invisible light shining through a wall. The availability
of several decommissioned LHC superconducting magnet prototypes at CERN,
offers the opportunity of possible upgrades for the proposed experiments,
each of them improving the sensitivity by increasing the vacuum light pass
in the transverse magnetic field with the connection of additional magnets.

\begin{figure}[t]
  \centering
  \includegraphics[width=0.82\textwidth]{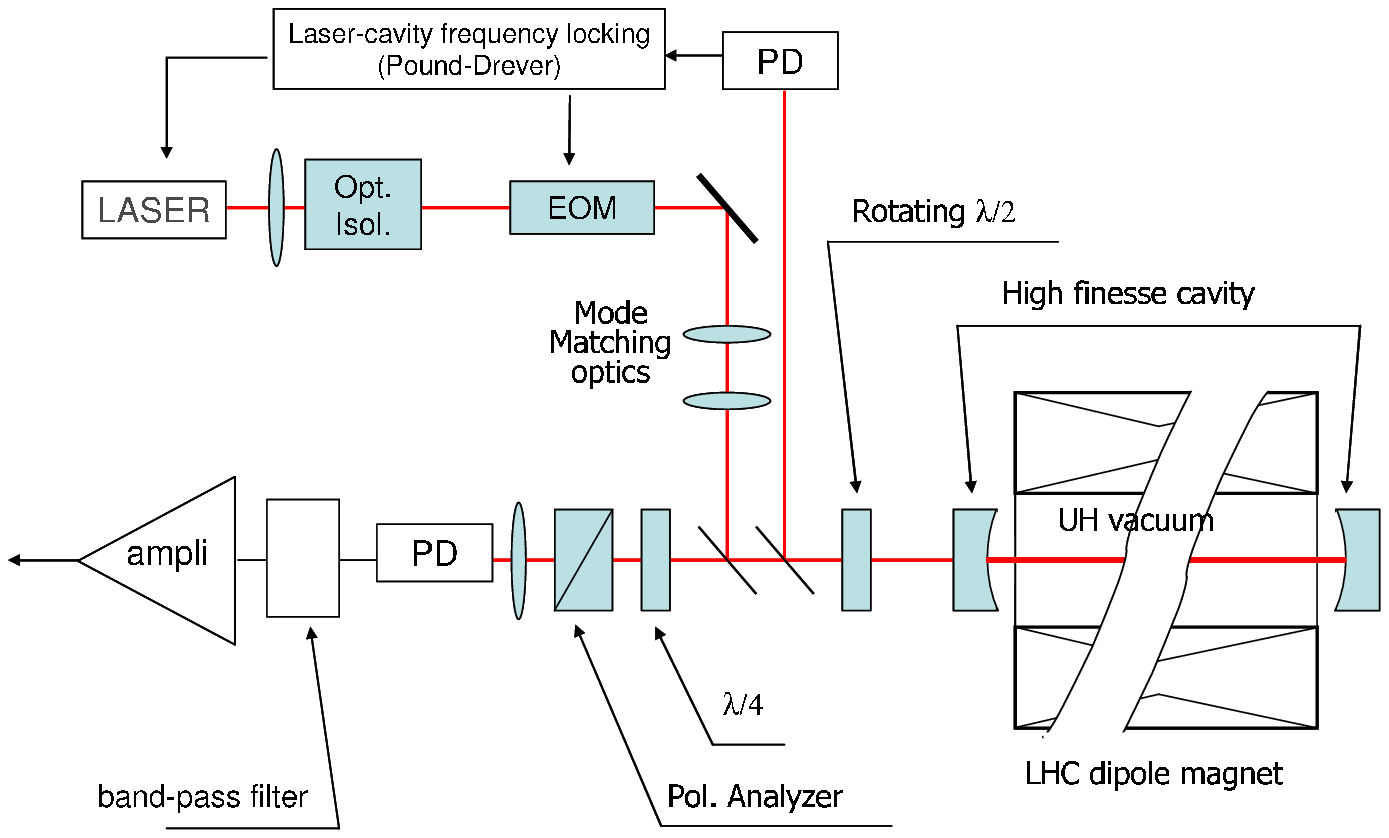}
  \caption{Optical scheme for the measurements of the vacuum magnetic
    birefringence adapted from \cite{ex-Pugnat} with inputs from D.
    Romanini and L. Duvillaret\label{fig:ex-pugnat}} 
\end{figure}

\paragraph{Measurement of the Vacuum Magnetic Birefringence (VMB)}
A first version of the experimental configuration, to measure very small
optical birefringence, was proposed by \cite{ex-Pugnat} and is shown in
Fig.  \ref{fig:ex-pugnat}. It is based on a Fabry-P\'{e}rot cavity of
finesse equal to $10^{3}$--$10^{4}$ and a novel measurement method, using a
double path of light through a half-wave plate, mounted in a high-speed
rotation stage.  The initial linear polarization state of the laser beam
can be modulated typically in the $\text{kHz}$ range. After the second pass
through the half-wave plate, the laser beam will retrieve to the first
order, its linear polarization state with the small VMB rotation angle,
induced by the vacuum and submitted to the transverse $B$ field inside the
optical cavity.  Then a quarter-wave plate will convert the quasi-linear
polarized laser beam coming out from the cavity into a quasi-circular one.
Finally, a polarizer will ensure a linear and optimal conversion of the
induced ellipticity into a power modulation of the laser beam. By replacing
the half-wave rotating plate with an electro-optic modulator, it can be
expected to work in the $\text{MHz}$ range for the modulation and the
detection.  This constitutes one of the alternative solutions studied at
present to improve the VMB measurements. For the phase-1 of this project,
the sensitivity is expected to reach the state of the art, i.e.,
$10^{-9}-10^{-8}\,\text{rad}/\sqrt{\text{Hz}}$. An improvement of at least
two orders of magnitude of the present reference results given by the BFRT
collaboration \cite{ex-BFRT}, concerning the VMB and absorption anisotropy,
is expected together with the measurement for the first time of the QED
prediction \cite{ex-Pugnat}.

\begin{figure}[t]
  \centering
  \includegraphics[width=0.8\textwidth]{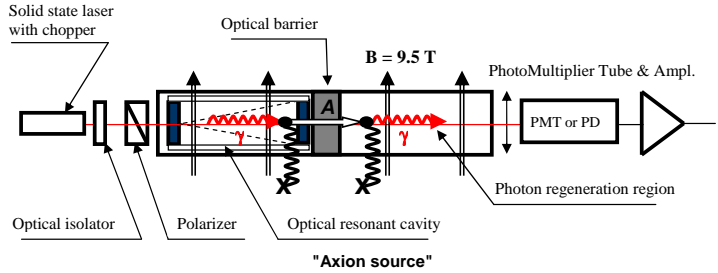}
  \caption{Scheme of the photon regeneration
    experiments\label{fig:ex-pungat-principle}} 
\end{figure}

\paragraph{The Photon Regeneration Experiment}
To complement and cross-check the VMB measurements, the photon regeneration
experiment will be integrated in the second aperture of the $15\,\text{m}$
long LHC superconducting magnet. The principle of this experiment is
schematized in Fig. \ref{fig:ex-pungat-principle}. A high finesse
Fabry-P\'{e}rot cavity, inserted inside a part of the dipole magnetic field
region, is used as an axion source and a photomultiplier with a proper
magnetic shield, or an avalanche photodiode, as an optical detection
system. The optical barrier will intercept all photons not converted into
axions and the detection of any photon at the same wavelength as the laser
beam can be interpreted as an axion to photon reconversion inside the
regeneration region. A chopper can be used for a synchronous photon
counting with the chopped laser beam, to improve the background rejection.
When the magnetic field is switched off, the same measurements can be
repeated to detect the possible mixing effect between photons and
paraphotons.

With a Nd:YAG laser, an optical beam power as large as
$100$--$1000\,\text{W}$ can be obtained at the wavelength $\lambda =
1064\,\text{nm}$. The optimum for the photon regeneration experiment is
obtained for an optical cavity and a regeneration region both immersed in
the same magnetic field integral.  Assuming as a first step the use of a
single LHC dipole with a $7\,\text{m}$ long regeneration region and an
optical cavity with a finesse of $10^{4}$--$10^{5}$ of the same length, the
photon counting rate given by the BFRT collaboration \cite{ex-BFRT} can be
improved by a factor of about $10^{8}$.  This corresponds to a limit for
the coupling constant to two photons $1/M$ of about
$10^{-9}\,\text{GeV}^{-1}$, i.e., an improvement by more than $2$ orders of
magnitude with respect to the BFRT results. The lost of coherence in the
axion to photon conversion will prevent to probe axions with a mass
typically larger than $0.4\,\text{meV}$ at the lower value of $1/M$.
During the preparatory phase of this photon regeneration experiment, the
focus will be given to the check of PVLAS results \cite{ex-PVLAS}.
Preliminary estimates show, that this objective can be achieved at $95\%$
confidence level with an integration time of about $1$ hour, assuming
$1000$ reflections in a cavity of $1\,\text{m}$ long in a $9\,\text{T}$
field and an optical power of about $1\,\text{W}$.

\section{Search for Kaluza-Klein Axions with TPCs}
\index{Axion!Kaluza-Klein!Search|\\see KK-Axion search} \index{KK-Axion
  search|(} Higher dimensional axionic theories that include $\delta$ extra
spatial dimensions to the $(3+1)$ dimensions of the Minkowski space-time
predict, that the axion field can propagate in these additional dimensions
(see Chap. \ref{chap:lacic-axions-and-large-extra-dimensions} for an
introduction).  An important consequence of such so called Kaluza-Klein
(KK) axion models would be, that the axion field would decouple from the
Peccei-Quinn energy scale and aquire an infinite tower of mass eigenstates.
The mass spacing between individual eigenstates would then be $\propto
1/R$, where $R$ is the compactification radius of the extra dimensions. For
the corresponding axion mass eigenvalues then follows
\begin{equation}
  \label{eq:ex-kk-axion-mass}
  m_{a0} \approx m_\text{PQ}\;\;\;\text{and} \;\;\; m_{an} \approx
  \frac{n}{R}\;\;\;\text{with}\;\;\; n=1,2,3, \ldots \;.
\end{equation}
Taking actual limits on the Peccei-Quinn scale into account, the lifetime
of the Peccei-Quinn axion with respect to a $a\rightarrow\gamma\gamma$
decay calculates from
\begin{equation}
  \label{eq:ex-kk-lifetime}
  \tau_a = \frac{64\pi}{g_{a\gamma\gamma}^2 m_a^3}\;,
\end{equation}
to $10^{27}\,\text{years} \lesssim \tau_a \lesssim 10^{42}\,\text{years}$,
which is several orders of magnitude longer compared to the age of our
universe ($13.7\times10^9\,\text{years}$ \cite{ex-Spergel:2003cb}).
Particles which decay with such long lifetimes are not observable with
actual experimental techniques. Taking \eqref{eq:ex-kk-axion-mass} and
\eqref{eq:ex-kk-lifetime} into account and setting $m_a = m_{a0}$, it is
obvious that the lifetime of the more massive KK-axion states is
significantly shorter, e.g., we get a lifetime of
$2.7\times10^{7}\,\text{years} \lesssim \tau_a \lesssim 2.7\times
10^{14}\,\text{years}$ for the $10\,\text{keV}$ mode. In this case, the
experimental observation of the axion di-photon decay becomes feasible, as
will be shown in the following sections.
\begin{figure}[t]
  \centerline{\includegraphics[width=0.70\textwidth]{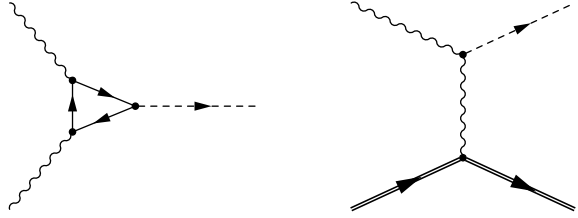}}
  \caption{Feynman diagram for axion production by photon coalescence
    (left) and the Primakoff effect (right)}
  \label{fig:ex-kk-axion-feynman}
\end{figure}

\subsection{Gravitationally Trapped Solar KK-Axions}
\begin{figure}[t]
  \centering
  \includegraphics[width=0.8\textwidth]{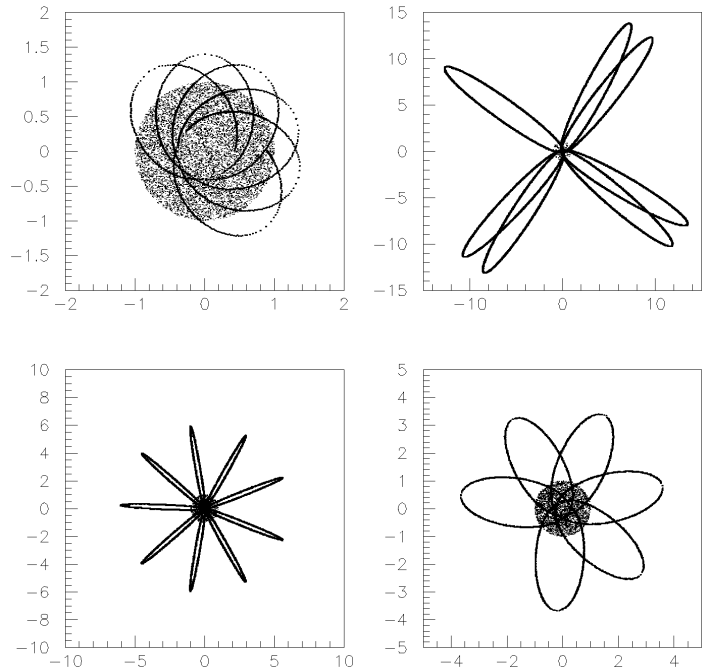}
  \caption{Keplerian orbits of Kaluza-Klein axions trapped in the
    gravitational potential of the Sun. The x- and y-axis indicate the
    distance from the Sun in units of solar radii (adapted from
    \cite{ex-dilella:03a} with permission)}
  \label{fig:ex-kk-axion-orbits}
\end{figure}

\begin{figure}[t]
  \begin{minipage}{0.49\textwidth}
    \centerline{\includegraphics[width=0.98\textwidth]{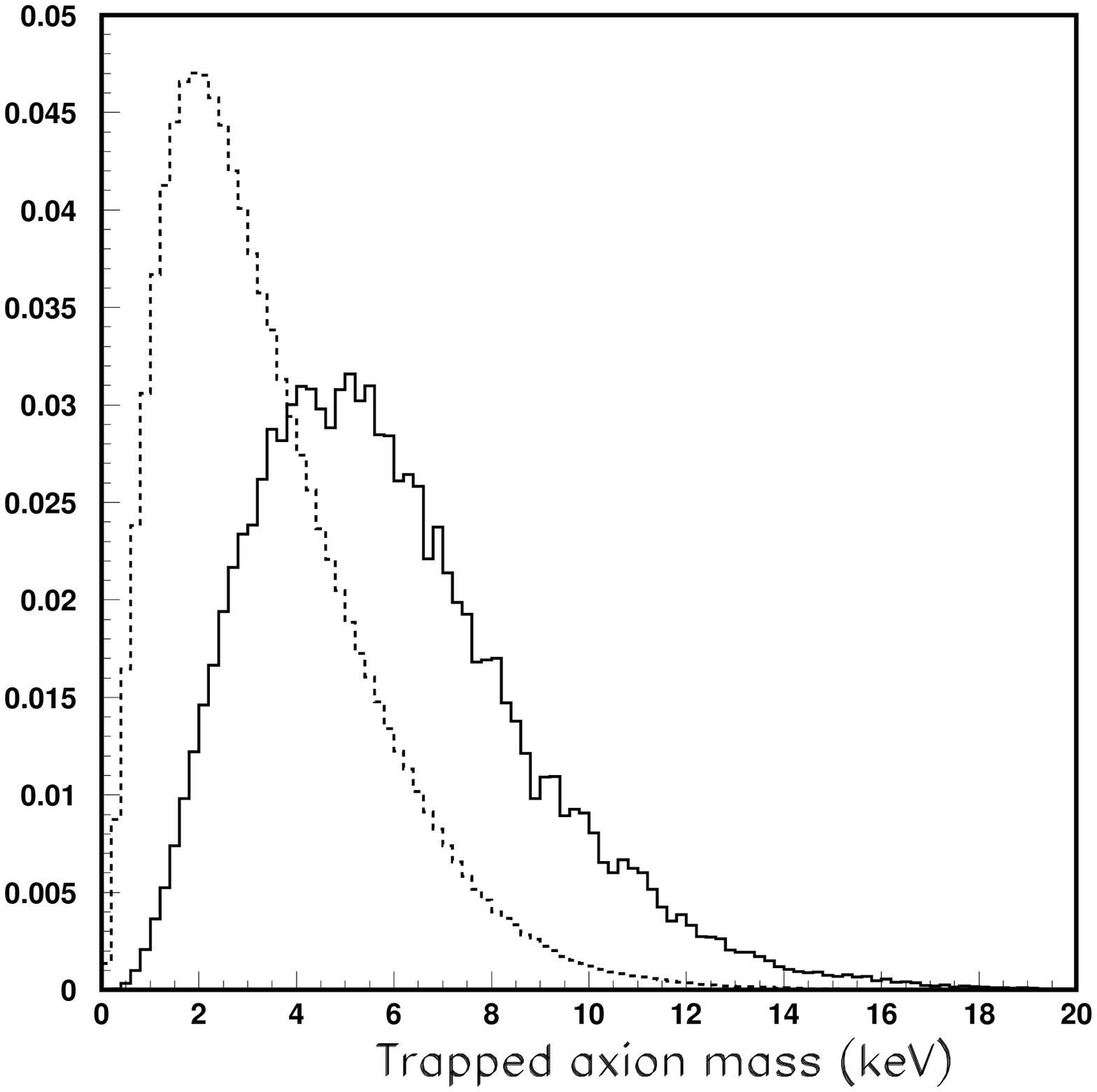}}
  \end{minipage}
  \begin{minipage}{0.49\textwidth}
    \centerline{\includegraphics[width=0.98\textwidth]{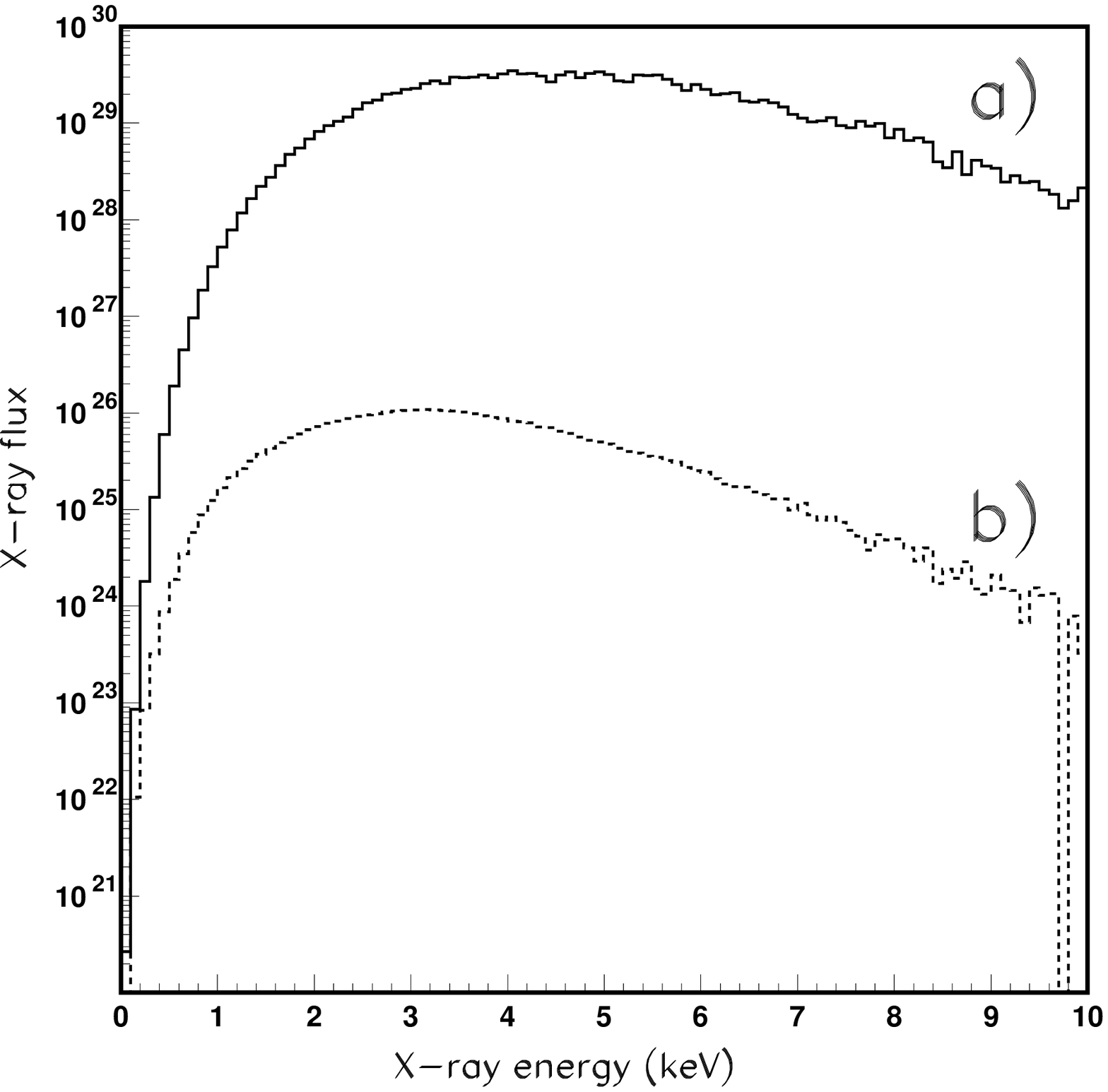}}
  \end{minipage}
  \caption{Left: Mass spectrum of KK-axions trapped in the gravitational
    potential of the Sun which cross the orbit of the Earth. The solid line
    represents the mass distribution for axions produced via photon
    coalescence and the dashed line the mass distribution for axions
    produced via the Primakoff effect. Please note that the area of both
    distributions is normalized to unity. Right: The corresponding X-ray
    spectrum resulting from axion to two photon decay of axions produced via
    a) photon coalescence and b) the Primakoff effect (adapted from
    \cite{ex-dilella:03a})}
  \label{fig:ex-kk-axion-spectra}
\end{figure}

\index{KK-Axion search!Gravitationally trapped|(}
A potential source of KK-axion would be the hot plasma of the Sun, were
axions can be produced by either the Primakoff conversion or photon
coalescence (see Fig. \ref{fig:ex-kk-axion-feynman}). The total solar axion
luminosity from both contributions is given by \cite{ex-dilella:03a}
\begin{equation}
  L_a = A L_\odot \left(
  \frac{g_{a\gamma\gamma}}{10^{-10}\,\text{GeV}^{-1}}\right)^2
  \left(\frac{R}{\text{keV}^{-1}} \right)^\delta\;,
\end{equation}
with a normalization factor $A$ ($A= 0.067/0.12$ for axions produced by
photon coalescence/Primakoff effect), the solar luminosity $L_\odot$, and
the number of extra dimensions $\delta$. Most of these axions would freely
stream out of the Sun and leave the solar system before they decay.
Nevertheless, a small fraction of axions would have velocities smaller than
the escape velocity $v_\text{esc}$ and could be trapped in the
gravitational potential of the Sun. These trapped KK-axions would revolve
the Sun on Keplerian orbits as shown in Fig.  \ref{fig:ex-kk-axion-orbits}
until they decay. Since axions produced by the Primakoff effect are
relativistic ($v \gg v_\text{esc}$), their contribution to the overall
number density of trapped axions is by a factor of $10^3$ smaller compared
to axions produced by photon coalescence. Depending on the KK-axion
lifetime $\tau_a$ the number density of trapped axions would evolve with
time according to \cite{ex-dilella:03a}
\begin{equation}
  N_a(t) = R_a\tau_a(1-\exp^{-t/\tau_a})\;,
  \label{eq:ex-number-density}
\end{equation}
where $R_a$ is the axion production rate. From \eqref{eq:ex-number-density}
the present KK-axion number density can be calculated depending on the
distance to the Sun, as well as the axion number density we would expect on
Earth today. The resulting mass spectrum of trapped KK-axions is shown in
the left part of Fig.  \ref{fig:ex-kk-axion-spectra} for both production
processes.  Assuming that KK-axions decay into two coincident photons with
identical energy $E_\gamma=m_a/2$, the decay spectrum is given by
\cite{ex-morgan:05a}
\begin{equation}
  \frac{\D R }{\D m_a} = \frac{g_{a\gamma\gamma}^2}{64\pi}n_0 m_a^3 f(m_a)
  \;,
  \label{eq:ex-decay-rate}
\end{equation}
where $f(m_a)$ is the trapped axion mass spectrum and $n_0$ the present
total KK-axion number density. The right part of Fig.
\ref{fig:ex-kk-axion-spectra} shows the axion decay spectrum resulting from
\eqref{eq:ex-decay-rate} on Earth depending on energy.  Integrating over
all energies, yields a total KK-axion decay rate of
\begin{equation}
  R = 2.5\times 10^{11} \left(\frac{g_{a\gamma\gamma}}{\text{GeV}^{-1}}
  \right)^2 \left( \frac{n_0}{\text{m}^{-3}}
  \right)\,\text{m}^{-3}\,\text{day}^{-1}\;. 
  \label{eq:ex-int-decay-rate}
\end{equation}
\index{KK-Axion search!Gravitationally trapped|)}

\subsection{KK-Axion Detection with Large TPCs}
\begin{figure}[t]
  \centerline{\includegraphics[width=0.80\textwidth]{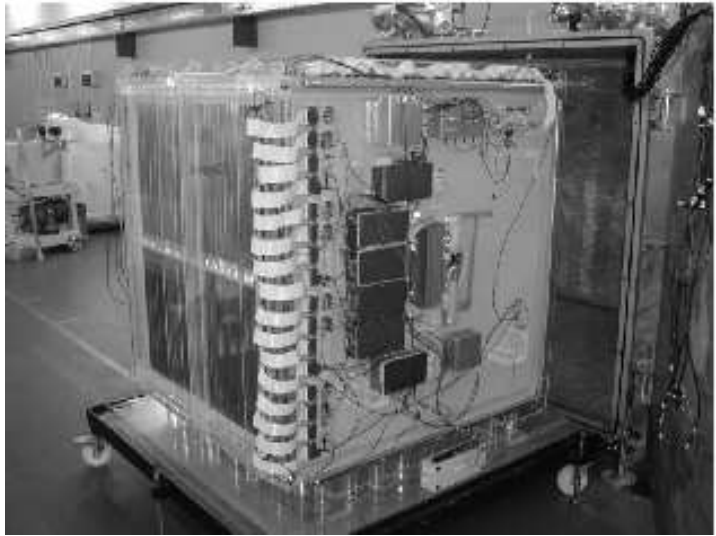}}
  \caption{Image of the DRIFT~II detector during installation in the Boulby
    underground laboratory}
  \label{fig:ex-drift-detector}
\end{figure}

\index{KK-Axion search!DRIFT|(} The possibility that such a local axion
population exists, motivates an experimental search for KK-axions. To
estimate the minimum sensitivity necessary to be able to observe KK-axion
decays, the expected axion di-photon decay rate can be calculated from
\eqref{eq:ex-int-decay-rate}.  Assuming a local axion number density of
$n_0 = 10^{14}\,\text{m}^{-3}$ and $g_{a\gamma\gamma} = 9.2\times
10^{-14}\,\text{GeV}^{-1}$ (see \cite{ex-morgan:04a,ex-dilella:03a} for
details), an axion decay rate of $R = 0.21\,\text{m}^{-3}\,\text{day}^{-1}$
is expected. It is obvious from this estimate that a detector with a large
sensitive volume and low background is required in order to be sensitive to
the resulting low count rates. In addition the detector must be able to
separate the two coincident axion decay photons with an energy of
$E_\gamma=m_a/2\approx \text{few}\,\text{keV}$ which are emitted back to
back.
\begin{figure}[t]
  \centerline{\includegraphics[width=0.70\textwidth]{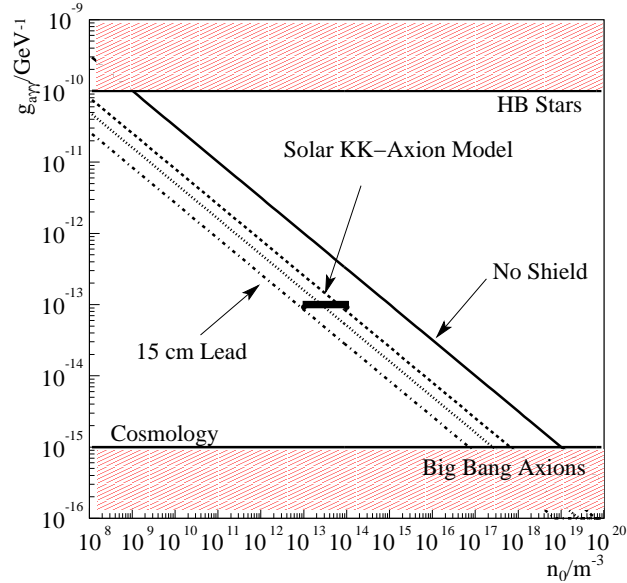}}
  \caption{Sensitivity estimate for a $1\,\text{m}^3$ TPC detector operated
    with CS$_2$ as gas for different detector shield configuration. In
    addition exclusion contours from astrophysical considerations are
    shown. The parameter range predicted from the solar KK-axion model of
    \cite{ex-dilella:03a} is marked by the black box (adopted from
    \cite{ex-morgan:05a})}
  \label{fig:ex-kk-exclusion-plot}
\end{figure}
Solid and liquid state detectors like NaI, Ge, or Xe are disfavored for
this purpose. The mean free path of X-rays in, e.g., Ge is of the order of
a few $\umu m$, consequently the decay photons would be indistinguishable
from the detector background. Instead, a large volume Time Projection
Chamber (TPC) operated underground and at low pressure similar to the
DRIFT~II\abbrev{DRIFT}{Directional Recoil Identification From Tracks,
  Boulby Underground Laboratory, UK}\index{DRIFT} detectors
\cite{ex-alner:05a} would be an ideal system for such an experiment. In
\cite{ex-morgan:05a} the sensitivity and background rejection efficiency
has been estimated based on Monte-Carlo simulations for a $1\,\text{m}^{3}$
DRIFT~II type detector operated with CS$_2$ as detection gas. The resulting
sensitivity is shown in Fig.  \ref{fig:ex-kk-exclusion-plot} for realistic
KK-axion model parameters and for different detector configurations.
According to these results, a DRIFT~II type TPC detector shielded with
$15\,\text{cm}$ Pb would be sufficient, to fully probe the predicted
parameter range of the model presented by \cite{ex-dilella:03a} (the range
of the model parameters is indicated as a black box in Fig.
\ref{fig:ex-kk-exclusion-plot}). At present it seems feasible to realize a
detector system at moderate cost by upgrading a DRIFT~II detector similar
to the one shown in Fig.  \ref{fig:ex-drift-detector}. This would require
additional shielding and a better spatial resolution of the detector in
order to reduce the detector background and to be able to spatially resolve
the secondary electrons produced by the two KK-axion decay photons.
\index{KK-Axion search|)}
\index{KK-Axion search!DRIFT|)}

\section{Collider Bounds on Scalars and Pseudoscalars}
If pseudoscalar particles couple to photons and gluons, it is possible to
constraint their couplings and mass by looking at processes where a
pseudoscalar particle is involved. The processes we consider here are
$e^+e^- \rightarrow \gamma +$ \MET\ for $e^+e^-$ colliders, and $pp$ or
$p\bar{p} \rightarrow$ single jet + \MET\ for hadron colliders. One
characteristic of the dimension five amplitudes considered here is that the
cross section is independent of the center of mass energy at high energies.
This follows from a dimensional analysis: the interaction $g \phi F \wedge
F$, where $\phi$ is the pseudoscalar field, has a coupling constant $g$
with a dimension of inverse mass. Therefore $2\rightarrow2$ processes,
involving the production of a single $\phi$ particle, will have a cross
section
\begin{equation}
  \frac{\D\sigma}{\D \Omega} = g^2 f(s/t)\,,
\end{equation}
where $f(s/t)$ is a function which depends only on the angle.
Specifically, the couplings we are interested are the following
\begin{equation}
  \label{eq:ex-glue}
  L \propto \frac{\alpha_s}{16 \pi f} \phi\,G_{\mu\nu}^b\tilde G^{b\mu\nu}\;, 
\end{equation}
for gluons, and 
\begin{equation}
  \label{eq:ex-light}
  L \propto \frac{g_{a\gamma\gamma}}{8} \phi\,F_{\mu\nu}\tilde F^{\mu\nu}\;,
\end{equation}
for photons.  Hadron colliders are more effective in providing limits for
axion-gluon couplings and $e^+ e^-$ colliders for axion-photon couplings.

\subsection{Bounds from Hadron Colliders}
\label{sec:ex-hadron-colliders}
Here the results for a two experiments in hadron colliders are summarized,
for a more detailed analysis we refer to \cite{ex-Kleban:2005rj}. Using the
data from the D\O\ experiment at Fermilab a bound on $f$ can be set to
\begin{equation}
  f > 35 \,\text{GeV}\;.
\end{equation}
As for future and ongoing experiments, the Large Hadron Collider (LHC) at
CERN is expected to improve the bound on $f$ to
\begin{equation}
  f_{\rm LHC} > 1300 \  {\rm GeV}\;.
\end{equation}

\begin{figure}[t]
  \centerline{\includegraphics[width=0.80\textwidth]{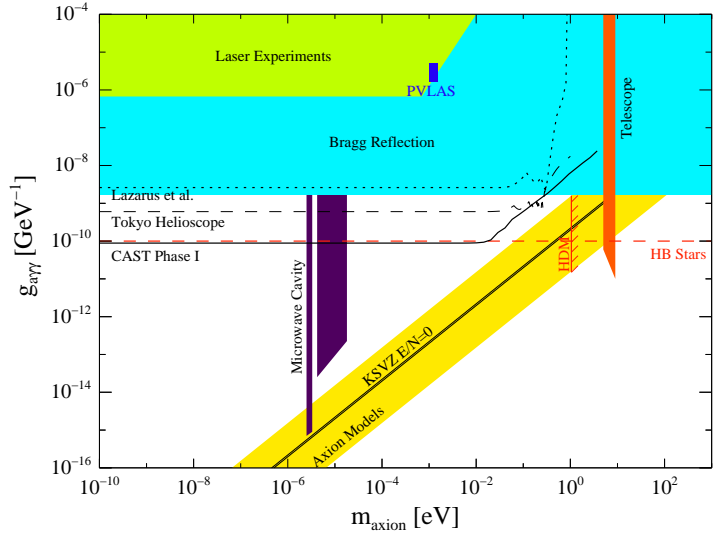}}
  \caption{Exclusion plots in the  $g_{a\gamma\gamma}$ versus $m_a$
    parameter space. Limits derived with different experimental techniques
    are shown. Limits from DAMA, SOLAX, and COSME are denoted as ``Bragg
    Reflection''
    \cite{ex-Morales:2001we,ex-Avignone:1997th,ex-Bernabei:2001ny}. In
    addition, the best limits from laser experiments, microwave cavity
    experiments, telescope searches, helioscope searches marked as
    ``Lazarus et al.'', ``Tokyo Helioscope'', and CAST
    \cite{ex-Andriamonje:2007ew,ex-Lazarus:1992ry,ex-Moriyama:1998kd,ex-Inoue:2000bj,ex-inoue:2002qy},
    and the best astrophysical limits are shown
    \cite{ex-Raffelt:2006cw,ex-Raf96}. The region predicted by theoretical
    models is marked as ``Axion Models'' ($E/N - 1.95$). The vertical line
    ``HDM'' indicates the hot dark matter limit for hadronic axions
    \cite{ex-hannestad:07a}}
  \label{fig:ex-exclusion-plot}
\end{figure}
\subsection{Bounds from $e^+ e^-$ Colliders}
\label{sec:ex-e-colliders}
More interesting are the limits that can be obtained for the photon
coupling to axions in $e^+ e^-$ colliders. Several experiments during the
past and future have been analyzed to determine the bound which could be
obtained assuming the data are consistent with the standard model
backgrounds. A bound $g_{a\gamma\gamma} < 5.5 \times
10^{-4}\,\text{GeV}^{-1}$ was obtained using $e^+e^-$ collider data from
ASP \cite{ex-Masso:1995tw} \cite{ex-Hearty:1989pq}.  Since the amplitudes
are independent of energy, the bounds can be improved mainly by increasing
the total luminosity. An analysis of the combined data from LEP at ALEPH,
OPAL, L3, and DELPHI would yield the more restrictive bound:
\begin{equation}
  g_{a\gamma\gamma} < 1.5 \times 10^{-4}\,{\rm GeV}^{-1}\;,
\end{equation}
for $m_\phi < 65\,\text{MeV}$.  For the PEP-II $e^+e^-$ collider, current
integrated luminosity gives
\begin{equation}
  g_{a\gamma\gamma}(\text{PEPII}) < 8.9 \times 10^{-6}\,{\rm GeV}^{-1}\;,
\end{equation}
for $m_\phi < 0.12\,\text{GeV}$. A similar bound can be obtained from KEKB
$e^+e^-$ collider. With the current integrated luminosity the bound would
be
\begin{equation}
  g_{a\gamma\gamma}(\text{KEKB}) < 8.2 \times 10^{-6}\,\text{\rm GeV}^{-1}\;,
\end{equation}
for $m_\phi < 0.13\,\text{GeV}$.  The expected total luminosity for KEKB is
at least twice the current total, which would improve the bound to
$g_{a\gamma\gamma} < 5.9 \times10^{-6}\,\text{GeV}^{-1}$. Finally, for the
Super KEKB upgrade to KEKB is expected to produce $10^7\,\text{pb}^{-1}$
per year.  After two years, this would improve the bound to
\begin{equation}
  g_{a\gamma\gamma}(\text{KEKB}) < 1.9\times10^{-6}\,\text{\rm GeV}^{-1}\;,
\end{equation}
which would rule out most of the parameter space favored by PVLAS.

\section{Summary and Outlook}
In this paper we reviewed different experimental searches for axions and
axion-like particles which are actually running or planned to take date in
the near future. Already existing upper limits on the axion to photon
coupling strength $g_{a\gamma\gamma}$ in the mass range of
$1\times10^{-10}\,\text{eV} \lesssim m_a \lesssim 1\times
10^{2}\,\text{eV}$ derived with different experimental techniques are
summarized in Fig. \ref{fig:ex-exclusion-plot}.  In addition the parameter
range that can be derived from the PVLAS result is indicated by a
rectangular box marked as ``PVLAS''.

It was not our intention to give a complete review of all existing
experimental techniques to search for axions, this would be beyond the
scope of an article like ours. Instead, we focused on the topics which were
addressed during the Joint ILIAS-CAST-CERN Axion Training at CERN. The
common factor of all experiments is the inspired techniques used, which are
extremely challenging and innovative.

The fact that the PVLAS collaboration has reported a positive signal that
could be interpreted as a signature of an axion-like particle has boosted a
race, to build a new set of laser-based experiments. We are convinced that
the following years are going to be very exciting.

\section{Acknowledgment}
We acknowledge support from the Virtuelles Institut f\"ur Dunkle Materie
und Neutrinophysik -- VIDMAN (Germany).  Furthermore, the authors acknowledge
the helpful discussions within the network on direct dark matter detection
of the ILIAS integrating activity of the European Union (contract number:
RII3-CT-2003-506222). Parts of this work has been performed in the CAST
collaboration, we thank our colleagues for their support. The CAST project
was supported by the Bundesministerium f\"ur Bildung und Forschung (BMBF)
under the grant number 05 CC2EEA/9 and 05 CC1RD1/0.

\input{experimentref}

\end{document}

%% file: experimentref.tex
%%% Local Variables: 
%%% mode: latex
%%% TeX-master: "axion-book"
%%% End: 